\documentstyle[12pt,epsfig]{article}

\renewcommand{\baselinestretch}{1.1}
\renewcommand{\arraystretch}{1.1}
\newcommand{\slh}{\hspace{-.5em}/}

\newcommand{\npb}[3]{Nucl.~Phys.~#1 (19#2) #3}
\newcommand{\prl}[3]{Phys.~Rev.~Lett.~#1 (19#2) #3}

\newcommand{\pr}[3]{Phys.~Rev.~D#1 (19#2) #3}

\newcommand{\lsim}{\raisebox{-0.13cm}{~\shortstack{$<$ \\[-0.07cm] $\sim$}}~}

\newcommand{\ra}{\rightarrow}
\newcommand{\ee}{e^+e^-}
\newcommand{\s}{\\ \vspace*{-3mm} }
\newcommand{\nn}{\noindent}
\newcommand{\non}{\nonumber}
\newcommand{\beq}{\begin{eqnarray}}
\newcommand{\eeq}{\end{eqnarray}}

\newcommand{\tb}{\tan\beta}

\begin{document}

\begin{titlepage}

\begin{flushright}
KA--TP--21--96\\
September 1996 \\
\end{flushright}

\def\thefootnote{\fnsymbol{footnote}}

\vspace{1cm}

\begin{center}

{\large\sc {\bf Associated Production of Higgs Bosons and a Photon}}

\vspace{.3cm}

{\large\sc {\bf in High--Energy $\ee$ Collisions}} 
\vspace{1cm}

{\sc A.~Djouadi\footnote{Supported by Deutsche Forschungsgemeinschaft
DFG (Bonn).}, V. Driesen, W. Hollik and J. Rosiek\footnote{Permanent 
adress: Institute for Theoretical Physics, Warsaw University, PL--00681 
Warsaw, Poland. Supported by the Alexander von Humboldt Stiftung
and by the Polish Committee of Scientific Research.}}

\vspace{1cm}

Institut f\"ur Theoretische Physik, Universit\"at Karlsruhe,\\
\vspace{0.2cm}
D--76128 Karlsruhe, FR Germany. \\

\end{center}

\vspace{1.6cm}

\begin{abstract}

We calculate the cross sections for the production of Higgs particles in
association with a photon in $\ee$ collisions, $\ee \ra \gamma+$Higgs,
allowing for the longitudinal polarization of the initial electron and
positron beams. We consider the associated production of both the
Standard Model Higgs boson, and the neutral CP--even and CP--odd Higgs
particles of its minimal supersymmetric extension. Complete and compact
analytical expressions are given, and the size of the cross sections is
illustrated for energies which will be reached at future $\ee$
colliders. 

\end{abstract}

\end{titlepage}

\def\thefootnote{\arabic{footnote}}
\setcounter{footnote}{0}
\setcounter{page}{2}

\setcounter{equation}{0}
\renewcommand{\theequation}{1.\arabic{equation}}

\section*{1. Introduction}

The search for Higgs particles and the study of the electroweak symmetry
breaking mechanism \cite{R1} is one of the most important goals of
future high--energy colliders. Once Higgs particles are found, it will
be of utmost importance to make a detailed study of their basic
properties. In particular, the measurement of the couplings of the Higgs
bosons to the other fundamental particles will be a crucial test of the
nature of the Higgs bosons. In this respect, future $\ee$ colliders
\cite{R2} will play a major role: the clean environment in these
machines will allow a rather precise determination of these couplings. \s 

In $\ee$ collisions, the couplings of the Standard Model (SM) Higgs
boson $H^0$ to the massive gauge bosons $W$ and $Z$ can be precisely
measured in the main production channels: the bremsstrahlung process
$\ee \ra H^0 Z$ \cite{R3} and the $WW/ZZ$ fusion mechanisms $\ee \ra W^*
W^*/ Z^* Z^* \ra H^0 \bar{\nu}_e \nu_e  / H^0 \ee$ \cite{R4}. The Higgs
couplings to fermions can also be measured in some mass range: for Higgs
masses below the $WW$ threshold, the strength of the Higgs couplings to
$\tau$ leptons and charm quarks, relative to the dominant coupling to
bottom quarks can be determined \cite{R5}. The large Higgs Yukawa coupling
to heavy top quarks can be measured in the production process $\ee \ra
\bar{t} t H^0$ \cite{R6} or if the Higgs is heavy enough to decay into
top quarks, in the process $\ee \ra H^0 Z \ra \bar{t}t Z$ \cite{R7}.
Finally, the trilinear Higgs self--coupling can be accessed via the
double Higgs production processes $\ee \ra ZH^0 H^0$ and $\ee \ra W^*
W^* \ra \bar{\nu}_e \nu_e H^0 H^0$ \cite{R8}. \s 

Another set of important Higgs boson couplings consists of the $H^0 gg$,
$H^0 \gamma \gamma$ and $H^0 Z \gamma$ vertices. These couplings do not
occur at the tree level but are induced by loops of heavy particles.
Because of the Higgs interaction being proportional to the particle
masses, the particles do not decouple for very large masses, and these
vertices could therefore serve to count the number of particles which
couple to the Higgs boson. The $H^0 gg$ vertex \cite{R8a} can be
accessed through the decay $H^0 \ra gg$ in $\ee$ collisions for Higgs
bosons in the mass range below $\sim 140$ GeV, or in the fusion reaction
$gg \ra H^0$ which is the main production mechanism of Higgs particles
at LHC. The $H^0 \gamma \gamma$ and $H^0 Z \gamma$ couplings can be
accessed in the decays $H^0 \ra \gamma \gamma$ \cite{R9,R9a} and $H^0
\ra Z \gamma$ \cite{R9a,R10} which typically have branching ratios of
${\cal O}(10^{-3})$ and could be measured at the LHC. The $H^0 \gamma
\gamma$ width can also be determined directly by means of the $\gamma
\gamma$ fusion process [with the photons generated by Compton
back--scattering of laser light] at high--energy $\ee$ colliders
\cite{R11}, while the $H^0 Z \gamma$ coupling can be measured in the
rare decay of the neutral vector boson $Z \ra H^0 \gamma$
\cite{R9a,R12,R9b}, if $H^0$ is light enough. \s 

A precise determination of these couplings could help to distinguish
between the SM Higgs boson and Higgs particles predicted by some of its
extensions such a two--Higgs Doublet Model (2HDM), supersymmetric
theories (SUSY) \cite{R1} or some other type of New Physics \cite{R13}.
For instance, in the Minimal Supersymmetric extension of the Standard
Model (MSSM), the Higgs sector is enlarged to contain two doublets of
scalar fields leading to a spectrum of five Higgs particles: two
CP--even Higgs bosons $h$ and $H$, a CP--odd Higgs boson $A$ and two
charged Higgs particles $H^\pm$. In the decoupling regime \cite{R14},
where the Higgs bosons $H,A$ and $H^\pm$ are very heavy, the lightest
MSSM Higgs particle $h$ has exactly the properties of the SM Higgs
boson, except that its mass is restricted to be smaller than $M_h \lsim
140$ GeV \cite{R15,R15b}. If the SUSY particles are too heavy to be produced
directly at the colliders, the only way to discriminate between 
the SM Higgs boson $H^0$ and $h$ in the decoupling limit is to look at the Higgs
couplings to $Z\gamma$ and $\gamma \gamma$. While in the SM these
couplings are induced by heavy fermion and $W$ boson loops only, the
MSSM provides additional contributions from loops involving charginos
and sfermions \cite{R16,R17}. \s 

An alternative way to have access to these induced Higgs--$\gamma \gamma$ 
and Higgs--$Z\gamma$ couplings is to consider the process 
\beq
\ee \ra H^0 \gamma \ . 
\eeq
In the SM, this process proceeds through $s$--channel $\gamma^* \gamma
H^0$ and $Z^* \gamma H^0$ vertex diagrams, but additional $t$--channel
vertex and box diagrams involving $W$/neutrino and $Z$/electron exchange
also occur (Fig.~1a). This process has been discussed some time ago in
Ref.~\cite{R18} and more recently in Ref.~\cite{R19} for unpolarized
electron and positron beams. Since it is a higher--order process in the
electroweak coupling\footnote{A similar process is the loop induced
double production of Higgs particles, $\ee \ra \Phi \Phi$, which has
been discussed in the SM and the MSSM recently \cite{R20}.}, the cross
section is rather small, but the signal is very clean allowing for a
resonable hope to isolate these events \cite{R21} if enough luminosity
is collected at a future high--energy collider. \s 

In this paper, we rederive the cross section for the process eq.~(1.1)
in the Standard Model, but allowing for the longitudinal polarization of
the initial $e^+$ and $e^-$ beams. In addition, we consider the
associated production of a photon and a Higgs boson in supersymmetric 
theories. In the MSSM, the associated production of the
CP--even neutral Higgs particles 
\beq
\ee & \ra & h \gamma \non \\
\ee & \ra & H \gamma
\eeq
and the production of the CP--odd Higgs boson
\beq
\ee \ra A \gamma
\eeq
will receive additional contributions coming from SUSY loops. For
the CP--even Higgs bosons one has additional $s$--channel $\gamma,Z$ 
exchange vertex diagrams involving loops with charged Higgs bosons, 
charginos, squarks and sleptons as well as $t$--channel vertex and 
box diagrams involving chargino/sneutrino and neutralino/selectron
loops (Fig.~1b). For the pseudoscalar state $A$, because of CP invariance, 
the process is mediated only by $s$--channel vertex diagrams involving
chargino and fermion loops, as well as $t$--channel vertex and box
diagrams involving chargino/sneutrino and neutralino/selectron loops
(Fig.~1c). \s

The rest of the paper is organized as follows. In the next section, 
we analyze the associated Higgs--photon production in the Standard
Model. In section 3, the case of the neutral CP--even and CP--odd Higgs 
bosons of the minimal supersymmetric extension is discussed. Some 
additional technical material is given in a short Appendix.

\section*{2. Associated $H \gamma$ production in the SM} 

\setcounter{equation}{0}
\renewcommand{\theequation}{2.\arabic{equation}}

\subsection*{2.1 Decomposition into form factors}

In the Standard Model, the process $\ee \ra \gamma H^0$ is described to
lowest order by the Feynman diagrams depicted in Fig.~1a. There are
$s$--channel $\gamma$ and $Z$ exchange vertex diagrams involving virtual
$W$ and heavy fermion loops [mainly top and bottom quark loops, since
the Yukawa couplings to the light fermions are very small], as well as
$t$--channel vertex diagrams involving $W/$neutrino and $Z$/electron
exchanges [corrections to the $H^0 ee$ vertex] and $W$/neutrino and
$Z$/electron box diagrams. Additional contributions come from diagrams
where the Higgs boson is emitted from the virtual $Z$ line and with
$Z$--$\gamma$ mixing through fermion or $W$ boson loops; diagrams with
the mixing between the $Z$ boson (or the photon) and the Higgs boson
give negligible contributions. \s 

Allowing for the polarization of both the initial electron and positron
beams, and summing over the polarizations of the photon, the
differential cross section of the process can be written as 
\beq
\frac{{\rm d} \sigma }{{\rm d} \cos\theta} (e^+e^- \to \gamma H^0) &=&
\frac{1}{2s}\frac{s-M_H^2}{16\pi s} \ \sum_{\rm pol} |{\cal M}|^2
\eeq
where $\sqrt{s}$ is the c.m.~energy and $\theta$ the scattering angle 
of the photon. Neglecting the electron mass, the amplitude ${\cal M}$ 
can be decomposed into the following sum of amplitudes 
\begin{equation}
{\cal M} \;=\; \sum_{i=1,3}\;\;\sum_{v=+,-} \Lambda_i^v \; C_i^v
\end{equation}
where the form factors $C_{i}^{\pm}$ sum all diagrams of Fig.~1a:
\begin{equation}
C_{i}^{\pm} \;=\; C^{\gamma\;\pm}_i +  C^{Z\;\pm}_i +  
C^{e\;\pm}_i + C^{\rm box\;\pm}_i   \ . \label{formfactor} 
\end{equation}
$C^{\gamma}$ and $C^{Z}$ denote the contributions of the $\gamma$ and
$Z$ pole vertex diagrams,  $C^{e}$ the $t$--channel $H^0ee$ vertex 
corrections and $C^{\rm box}$ the contributions of the box diagrams; the contributions due
to the $Z$--$\gamma$ mixing are included in $C^{Z}$. Denoting by
$p_\pm$ the momentum [all momenta are taking to be incoming] of the
initial $e^\pm$ beams, $p_\gamma$ and $\epsilon_\gamma$ the photon 
momentum and polarization vector, the set of standard matrix elements
$\Lambda_i^{\pm}$ is given by 
\beq
\Lambda^{\pm}_{1} &=& \overline{v}(p_+)\; (1 \pm \gamma_5)
\big( \epsilon\slh_\gamma \; p_{\gamma}.p_{-}
       - p\slh_{\gamma} \; \epsilon_\gamma.p_{-} \big) \;u(p_-) \non\\
\Lambda^{\pm}_{2} &=& \overline{v}(p_+)\; (1 \pm \gamma_5)
 \big( \epsilon\slh_\gamma \; p_{\gamma}.p_{+}
       - p\slh_{\gamma} \; \epsilon_\gamma.p_{+} \big) \;u(p_-) \non\\
\Lambda^{\pm}_3 &=& \overline{v}(p_+)\; (1 \pm \gamma_5)
 \big( \epsilon\slh_\gamma \;\big)\; u(p_-) \ .
\eeq
Summing over the photon polarizations, the squared amplitude can be 
written as 
\beq
\sum_{\rm pol} |{\cal M}|^2 = \frac{1}{(16\pi^2)^2} 
\bigg[ \, \sum_{v=+,-} \ \sum_{i,j=1,3} 
\; C_i^v \; T_{ij}^{vv}\; (C_j^v)^* \, \bigg] 
\eeq
with 
\beq
T_{ij}^{++} \;=\; \sum_{\rm pol} \Lambda^+_i  (\Lambda^+_j)^{\dagger}
&=&  \frac{1}{4} (1+\lambda_-)(1-\lambda_+) \; T_{ij} \non \\
T_{ij}^{--} \;=\; \sum_{\rm pol} \Lambda^-_i  (\Lambda^-_j)^{\dagger} 
&=&  \frac{1}{4} (1-\lambda_-)(1+\lambda_+) \; T_{ij} \non
\eeq
\beq
T_{ij}^{-+} = T_{ij}^{+-}  &=& 0 
\eeq
where $\lambda_-$ and $\lambda_+$ are the longitudinal polarizations of
the initial $e^-$ and $e^+$ beams, respectively. In terms of the usual
Mandelstam variables $s=(p_-+p_+)^2$, $u=(p_-+p_{\gamma})^2$ and
$t=(p_++p_{\gamma})^2$, the matrix $T$ reads 
\beq
T_{ij} = 2s\; \left(
\begin{array}{ccc}
 u^2  &   0   &  2 u  \\
 0    &  t^2  &  2 t  \\
 2 u  &  2 t  &   4
\end{array} \right)  \ . 
\eeq

In fact, as we will discuss later in more details, QED gauge invariance
requires that the form factors $C^{\pm }_{3}$ vanish after summing over
the contributions of all Feynman diagrams. This is obvious from
eq.~(2.4): while the form factors $\Lambda_{1,2}^\pm$ vanish if the
polarization of the photon $\epsilon_\gamma$ is replaced by its momentum
$p_\gamma$, it is not the case for $\Lambda_3^\pm$; the transversality
of the photon therefore implies that it is the amplitude $C_3^\pm$ which
should vanish. Therefore, one can drop the contributions of 
$\Lambda_3^\pm$ and the differential cross section of the process $\ee 
\to\gamma H^0$ can be written in the simple form
\beq
\frac{{\rm d}\sigma}{{\rm d}\cos\theta} &=& 
\frac{s-M_H^2}{64 \pi s} \frac{1}{(16\pi^2)^2}  \bigg\{
 (1+\lambda_-)(1-\lambda_+) 
 \bigg[ u^2\; |C_1^+|^2 + t^2\; |C_2^+|^2 \bigg] \non\\
&& \mbox{}\hspace{2.5cm}
   + (1-\lambda_-)(1+\lambda_+) 
 \bigg[ u^2\; |C_1^-|^2 + t^2\; |C_2^-|^2 \bigg] \bigg\} \ , 
\eeq
where the scattering angle $\theta$ is related to $t$ by $t = -(s-M_H^2)
(1-\cos \theta)/2$. The unpolarized cross section is obtained by 
simply setting $\lambda_\pm=0$: 
\beq
\frac{{\rm d} \sigma^{\rm unpol.}}{{\rm d} \cos\theta} =
\frac{s-M_H^2} {64 s\pi} \frac{1}{(16\pi^2)^2}
\bigg[ u^2\; \left( |C_1^+|^2+|C_1^-|^2 \right) + t^2\;
\left(|C_2^+|^2+|C_2^-|^2 \right) \bigg] \ . 
\eeq

\subsection*{2.2 $s$--channel diagrams}

Let us now discuss the contributions of the various diagrams to the form
factors $C_i^\pm$ and start with the $s$--channel vertex and $Z$--photon
mixing contributions. The contributions of the fermions and $W$ bosons to 
the $\gamma\gamma H^0$ or $Z\gamma H^0$ vertex diagrams can be decomposed 
into the following tensorial form [all momenta are taken as incoming and 
$V \equiv Z,\gamma$]: 
\beq
\Gamma [V^{\mu}(p_V),\gamma^{\nu}(p_\gamma),H^0(p_H)] &=& 
G^V_1 g^{\mu\nu} 
+ G^V_2 p_\gamma^{\mu} p_\gamma^{\nu} 
+ G^V_3 p_\gamma^{\mu} p_V^{\nu}
+ G^V_4 p_V^{\mu} p_\gamma^{\nu} \non \\
&& + G^V_5 p_V^{\mu} p_V^{\nu} 
+ G^V_6 \epsilon^{\mu\nu\alpha\beta} {p_V}_{\alpha} {p_\gamma}_{\beta} 
\label{tdec}
\eeq
Neglecting the electron mass, for on--shell final photons only 
the form factors $G_1$ and $G_3$ contribute to the amplitude. They 
are related to the form factors $C_{1,2}^{Z, \gamma}$ by
\beq
C^{\gamma\;\pm}_1 \;=\; C^{\gamma\;\pm}_2 &=&
 - \frac{e}{2}\,\frac{1}{s}\; G_3^{\gamma} \non\\ 
C^{Z\;\pm}_1 \;=\; C^{Z\;\pm}_2 &=&
 -\frac{e\, z^{\pm}}{4 s_W c_W}\, \frac{1}{s-M_Z^2}\; G_3^{Z} \label{formfi}
\eeq
where $e$ is the electric charge, $ z^+ = -1 + 2 s_W^2$ and $z^-=2s_W^2$
with $s_W^2=1-c_W^2 \equiv \sin^2\theta_W$. Although only the form factors 
$C_{1,2}^{Z,\gamma}$ contribute to the $\ee \ra H^0\gamma$ cross section, we 
also give the expressions for $C_3^{Z,\gamma}$ to check explicitly that 
QED gauge invariance is indeed fulfilled: 
\beq
C^{\gamma\;\pm}_3 &=&
  \frac{e}{2}\, \frac{1}{s}\; 
\left[ G_1^{\gamma} - \frac{s-M_H^2}{2}\;G_3^{\gamma}\right] \non \\
C^{Z\;\pm}_3 &=&
  \frac{e\, z^{\pm}}{4 s_Wc_W}\,\frac{1}{s-M_Z^2} \; 
 \left[ G_1^{Z} - \frac{s-M_H^2}{2}\;G_3^{Z}\right] \ . 
\eeq
Note that in the previous expressions we have omitted the finite width 
in the $Z$ boson propagator for simplicity. \s

The form factors $G_i^{\gamma,Z}$ are obtained by summing bosonic and
fermionic contributions to the $V \gamma H^0$ vertices:
\beq
G_i^{\gamma} & = & 
\frac{e^3\;M_W}{s_W} \left[ \,F^{\gamma,W}_i 
-\sum_{f}\, 4Q_f^2\,N_c\,\frac{m_f^2}{M_W^2} \,F^f_i \right] \non\\
G_i^{Z} & = & \frac{e^3\,M_W}{c_W\,s_W^2} \left[
 \,F^{Z,W}_i + \sum_{f}\, 2Q_f\,N_c\, \frac{m_f^2}{M_W^2} 
(I^f_3-2s_W^2 Q_f)  \,F^f_i \right] \label{giz}
\eeq
where $Q_f$, $I^f_3$ and $m_f$ are the electric charge, weak isospin and 
mass of the fermion $f$ [in practice only top and bottom quarks] 
respectively, and $N_c$ is the color factor, $N_c=1$ for leptons and 
$N_c=3$ for quarks. \s

The explicit calculation of the form factors $G_{1, 3}^{\gamma,Z}$\, was
done in the Feynman gauge where the $W$ boson contributions can be split
into the $g^{\mu \nu}$ and charged Goldstone parts. The ultraviolet
divergent amplitudes have been reduced from a complicated tensorial form
to scalar Passarino-Veltman two-- and three--point functions \cite{S1}
using the package FORM \cite{S2}. These scalar functions are defined in
the Appendix and their complete analytical expressions can be found in
Ref.~\cite{S3} for instance; their numerical evaluation has been
performed using the package FF \cite{S4}. \s 

The fermion loops give the same contribution $F^f_{1,3}$ to the $\gamma^*
\gamma H^0$ and $Z^* \gamma H^0$ vertices since the difference due to
the different $\gamma ff$ and $Zff$ couplings has been factorized out in
eq.~(2.13): 
\beq
F^f_1 &=& \frac{1}{2} 
 \Big[ 1 + (2 m_f^2 - M_{H}^2 - s ) C_0 - 2 s(2 C_{11} + C_{21}) 
    + 2 ( s - M_{H}^2) (C_{12}+ C_{23})  \Big] \non \\
F^f_3 &=&   C_0 + 4 C_{12} + 4 C_{23} 
\eeq
with the $C_0$ and $C_{ij}$ three--point functions defined as
\beq
C_{ij} \equiv C_{ij}(s,0,M_H^2;m_f^2,m_f^2,m_f^2) \ . 
\eeq
Note that the fermionic contribution to the $Z$--$\gamma$ mixing
diagrams is proportional to the final photon momentum squared and
therefore vanishes for photons on the mass shell. \s

The $W$ boson loops involving true vertex diagrams [with the exchange of
$W$ bosons, charged Goldstones and ghosts] and two--point functions
[involving the quartic couplings between two scalars and two vectors]
give different contributions to the $\gamma^* \gamma H^0$ and $Z^*
\gamma H$ vertices because of the complicated structure of the trilinear
gauge boson and quartic gauge boson scalar couplings which are
different for the $Z$ and the photon. Summing over all diagrams, the $W$
boson contribution reads
\beq
F^{\gamma,W}_1 &=& 
 \left( \frac{M_H^2}{M_W^2} +6  \right)   ( 4 C_{24} - B_{13} )
 + M_H^2   (  - 7 C_0 + C_{11} + C_{12} ) \non \\
&&   + s (5 C_0 + C_{11} - C_{12} ) - B_{12} + B_{23}  \non \\
F^{\gamma,W}_3 &=&
 4\left( \frac{M_H^2}{M_W^2} +6 \right)  ( C_{12} + C_{23} ) + 16 C_0 
\eeq
for the $\gamma^* \gamma H^0$ vertex, and
\beq
F^{Z,W}_1 &=& 
 \frac{1}{2} \frac{M_H^2}{M_W^2}( 1 - 2 c_W^2) ( 4 C_{24} - B_{13} )
 + M_H^2 c_W^2    ( 7 C_0 - C_{11} - C_{12} - 2 C_{23} ) \non\\
&& + M_H^2   (  - C_0 + C_{11} - C_{23} )
 + 2 M_W^2 (c_W^2-1) C_0  \non\\
&& - c_W^2 \left[ s (5 C_0 + C_{11} - C_{12} + 2 C_{21}
   - 2 C_{23} ) + 32 C_{24} - B_{12} - 6 B_{13} - B_{23} \right] \non \\
&& + \frac{1}{2} + c_W^2 + s ( C_0 - C_{11} - C_{21} + C_{23} ) - B_{12} 
+ B_{23} \non \\
F^{Z,W}_3 &=& 2 \left[ \frac{M_H^2}{M_W^2} (1-2 c_W^2)   + 2 ( 1-6 c_W^2 ) 
\right] (  C_{12} + C_{23} ) + 4 ( 1 - 4 c_W^2 ) C_0 
\eeq
for the $Z^* \gamma H^0 $ vertex. The two--point and three--point 
functions, $B_{ij}$ [note that here, the $B_{ij}$ functions are 
different from the standard ones given in \cite{S1} for instance] 
$C_{ij}$ are defined as
\beq
B_{12} & \equiv & B_0(s;M_W^2,M_W^2) \non\\
B_{13} & \equiv & B_0(M_H^2;M_W^2,M_W^2) \non\\
B_{23} & \equiv & B_0(0;M_W^2,M_W^2) \non\\
C_{ij}   & \equiv & C_{ij}(s,0,M_H^2;M_W^2,M_W^2,M_W^2)
\eeq
The expressions in eq.~(2.17) include the $W$ boson contribution to the 
$Z$--$\gamma$ mixing diagrams, which is non--zero only in the case of the 
form factor $F_{1}^{Z,W}$ and reads 
\beq
F_1^{Z,W}|_{\rm mix}  =  2 B_{23}  =  2 B_0(0;M_W^2,M_W^2) \ . 
\eeq

\vspace*{2mm}

Finally, we note that while the fermionic contributions are gauge 
invariant by themselves, i.e. that one has $C_3^{\gamma}+C_3^{Z} 
\equiv 0$ for fermions, the $W$ contributions are not. Gauge invariance
of the $W$ contributions is obtained only when the $t$--channel 
vertex and the box diagrams are included. 

\subsection*{2.3 $t$--channel vertex and box diagrams}

The contributions of the $t$--channel $H^0\ee$ vertex corrections 
come from $W$/neutrino and $Z$/electron loops. Only the ``transverse"
part of the gauge bosons contribute since the longitudinal or Goldstone
part is proportional to the very small electron mass. The contribution
of these diagrams to the form factors $C_i^e$ can be written as
\beq
C^{e\;\pm}_i &=& \frac{e^4}{s_W^3}\left[ \frac{M_W}{2}  \,{A}_i^{W\pm} 
   +\frac{M_Z}{4\,c_W^3} \,{A}_i^{Z \pm} \right]  \ + \ {\rm crossed}
\eeq
with the $W$/neutrino contributions 
\beq
A_3^{W+} &=& C_{12}(m_e^2,t,M_H^2;M_W^2,0,M_W^2) \non \\
A_3^{W-} &=& 0 \non \\
A_1^{W\pm} &=& A_2^{W\pm}=0  
\eeq
and the $Z$/electron contributions
\beq
A_3^{Z\pm} &=& (z^\pm)^2 \, C_{12}(m_e^2,t,M_H^2;M_Z^2,m_e^2,M_Z^2) \non \\
A_1^{Z\pm} &=& A_2^{Z \pm} = 0 \ . 
\eeq
The crossed terms are obtained by substituting $t \ra u$ 
in the functions $C_{12}$ of eqs.~(2.21) and (2.22). 
Because of the $t$--channel
electron exchange, terms proportional to ln$(m_e^2)$ appear and
one therefore has to keep a finite value for the electron mass in the
argument of the $C_{12}$ functions. This dependence on ln($m_e)$ will
disappear after adding the $W$ and $Z$ box contributions. \s 

Finally, the contribution of the box $W$ and $Z$ diagrams to the form 
factors $C_i^{\rm box}$ can be written as
\beq
C^{\rm box \pm}_i =  -\frac{e^4 M_W}{4\,s_W^3} \, \left[  B^{W \pm}_i + 
{\rm crossed} \right] \, 
+ \, \frac{e^4 M_Z}{4 s_W^3 c_W^3} \, B_i^{Z \pm} \ . 
\eeq
The terms from the $W$/neutrino box contribution read:
\beq
B_1^{W+} &=&  4 \Big( D_0 + D_{11} - D_{13} + D_{23} - D_{25} \Big)\non\\
B_2^{W+} &=&  4 \Big(  - D_{12} + D_{13} + D_{23} - D_{26} \Big)\non\\
B_3^{W+} &=&   \frac{1}{2}
\Big[ 3 s (- D_{12}+ D_{13}+ D_{25}+ D_{26}- D_{23}- D_{24}) \non\\
&& \mbox{}\hspace{3mm}
 + 7 t (- D_{23}+ D_{26}) + u (3 D_{13}+7 D_{25}-7 D_{23}) - 20 D_{27} \Big]
\ . 
\eeq
Since the $W$ boson is left--handed, there is no contribution from the
$W$ boxes to the form factors $B_i^{W-}$:
\beq
B_{1,2,3}^{W-}=0 \ . 
\eeq
The four--point functions $D_0$ and $D_{ij}$ have to be understood as 
\beq
D_{ij}= D_{ij}(m_e^2,m_e^2,M_H^2,0,s,u;M_W^2,0,M_W^2,M_W^2)
\eeq
in the expressions given above, while in the crossed contributions 
one has to interchange the photon and Higgs boson masses and interchange 
$u \leftrightarrow t$ in eqs.~(2.24) and (2.26)
\beq
D_{ij}= D_{ij}(m_e^2,m_e^2,0,M_H^2,s,t;M_W^2,0,M_W^2,M_W^2) \ . 
\eeq

The $Z$/electron box contributions in eq.~(2.23) are given by
\beq
B_1^{Z \pm} &=& 2 (z^\pm)^2 \, ( - D_{11} + D_{12} + D_{22} - D_{24} ) \non \\
B_2^{Z \pm} &=& 2 (z^\pm)^2 \, ( D_{22} - D_{26} ) \non \\
B_3^{Z \pm} &=& (z^\pm)^2 \, 
 \left[  s ( D_{22} - D_{24} + D_{25} - D_{26} ) + 2 D_{27} 
\right] 
\eeq
where 
\beq
D_{ij}= D_{ij}(m_e^2,M_H^2,m_e^2,0,t,u;M_Z^2,m_e^2,m_e^2,M_Z^2) \ . 
\eeq
There is no crossed contribution here since the photon does not couple
to the $Z$ boson directly. Again, we have kept the electron mass in the
arguments of the functions $D_{ij}$ to avoid infrared singularities;
these logarithmic singularities cancel exactly, as it should be, those
which appear in the $t$--channel $H^0 ee$ vertex contributions. 

\subsection*{2.4 Results} 

We have verified explicitly that the sum of the $W$ and the $Z$ boson
contributions from the various diagrams is gauge invariant, i.e. that
the form factor $C_3$ in eq.~(2.3) indeed vanishes when all
contributions are added. This cancellation occurs as follows: the
$Z$/electron contributions to the $H^0ee$ vertex and the box diagrams
cancel each other, $C^{e\; \pm}_{3} + C^{\rm box\;\pm}_{3} = 0$.
Furthermore, the $W$ boson contributions to $C^{\gamma\;-}_{3}+
C^{Z\;-}_{3}$ is zero, and as discussed previously the contributions to
$C_{i}^{-}$ of the $W$ boson induced $H^0\ee$ vertex and box diagrams
are zero since the $W$ boson is left--handed. Finally, the form factor
$C_3^+$ vanishes after summing the $W$ boson contributions of all
diagrams. This feature, together with the fact that the sum of all
contributions is both ultraviolet and infrared finite, provides a very
strong check of the calculation. \s 

We have also compared our results with previous calculations of the $\ee
\ra H^0 \gamma$ cross section in the unpolarized case \cite{R18,R19}.
Our analytical results agree completely with the results\footnote{The
results of Ref.~\cite{R19} have been obtained using a non--linear gauge
and the individual contributions are therefore different from ours;
however the sum of all contributions is gauge invariant and can be
compared with our final result.} of Ref.~\cite{R19}, provided that their
fermionic contribution to the $s$--channel $\gamma^* \gamma H^0$ and $Z
\gamma H^0$ vertices is multiplied by a factor of two. Our result for
the fermionic contributions agrees with the one obtained in
Ref.~\cite{R9a,R9b,R12} for the $Z \ra H^0 \gamma$ decay and also with
Ref.~\cite{R18} if the color factor for quarks is included in the
fermionic sum. We have also compared numerically our results with those
of Ref.~\cite{R18}, who calculated the $\ee \ra H^0 \gamma$ cross
section using the Feynman gauge. Our numbers agree with those given in
Tab.~4 of Ref.~\cite{R18} up to a few percent; the small difference is
probably due to a slightly different choice of input parameter values.
\s 

The cross sections for the process $\ee \ra H^0 \gamma$ are shown in 
Fig.~2 as a function of the Higgs boson mass for two center of mass
energies $\sqrt{s}=500$ GeV and $1.5$ TeV typical of future high--energy 
$\ee$ colliders. At 500 GeV, the unpolarized cross section [solid curve] 
is of the order of $\sigma \sim 0.05$ fb for light Higgs masses and 
increases slightly when approaching the $WW$ threshold; it then drops 
out quickly with increasing $M_H$ due to the lack of phase space. 
At $\sqrt{s}=1.5$ TeV, the cross section for light $H^0$ drops by a factor 
of $\sim 5$ compared to the previous case, but the decrease with 
increasing $M_H$ is slower. \s

Fig.~2 shows also the effect on the cross section of polarizing
longitudinally the electron and positron beams. With left-handed
longitudinally polarized electrons [dashed lines] the cross section
$\sigma(e_L^- e^+ \ra H^0\gamma$) is approximately two times larger, and
with left-handed electrons and right--handed positrons [dotted lines]
the cross section $\sigma(e_L^- e^+_R \ra H^0\gamma$) is approximately
four times larger than in the unpolarized case. The longitudinal
polarization of both the electron and positron beams is therefore
important to increase the cross section; for energies in the 300 GeV
range the cross section can reach values of the order of 0.5 fb
for $M_H \sim 100$ GeV. \s 

Fig.~3a exhibits the dependence of the cross section on the center of
mass energy for several values of the Higgs boson mass. The cross
section increases rapidly with the opening of the phase space and then
decreases near $\sqrt{s} \sim 350$ GeV close to the $t\bar{t}$ threshold
[the $W$ and top quark contributions interfere destructively, and the
top contribution is maximal near the $t\bar{t}$ threshold]. The cross
section drops smoothly with increasing energy; despite of the
presence of $t$--channel vertex and box contributions, it 
scales approximately as $1/s$ at high energies.  For $M_H \sim 100$
GeV, the cross section can be maximized by running the collider at
energies around 250 GeV. \s 

Finally, Fig.~3b shows the angular distribution d$\sigma$/d$\cos \theta$
for a c.m. energy of 500 GeV and a Higgs boson mass $M_H=100$ GeV. The
distribution is forward--backward symmetric and does not depend very
strongly on the Higgs boson mass. \s 

With the yearly integrated luminosity of ${\cal L} \sim 100$ fb$^{-1}$
expected at future high--energy $\ee$ colliders, one could collect a few
tens of $H^0 \gamma$ events in the course of a few years, if
longitudinal polarization is available\footnote{Note that at the CERN
collider LEP2 with a c.m. energy of $\sqrt{s} \sim 180$ GeV, the
expected luminosity which will be available, ${\cal L} \sim 0.5$
fb$^{-1}$, does not allow to produce any $H^0 \gamma$ event.}. The
signal, which mainly consists of a photon and a $b\bar{b}$ pair in the
low Higgs mass range or $WW$ and $ZZ$ pairs for Higgs masses larger than
$\sim 160$ GeV, is extremely clean. The backgrounds should be very small
since the photon must be very energetic and the $bb$ or $WW/ZZ$ pairs
should peak at an invariant mass $M_H$ [these rare events will be
searched only once the Higgs boson is found in one of the main
production channels and the Higgs mass would be precisely known by
then]. Therefore, despite of the low rates, the clean signal gives a
reasonable hope to isolate these events.

\section*{3. Associated production of MSSM Higgs bosons} 

\setcounter{equation}{0}
\renewcommand{\theequation}{3.\arabic{equation}}

In the MSSM, the associated production of the CP--even Higgs bosons $h$,
$H$  and the CP--odd Higgs boson $A$ with a photon receives extra
contributions from loops involving charged Higgs bosons and genuine
supersymmetric particles. More precisely, the production of the CP--even
Higgs particles, $\ee \ra \Phi \gamma$ with $\Phi=h,H$, proceeds
through the same diagrams as for the SM Higgs boson substituting the 
Higgs couplings to $W,Z$ bosons and fermions, plus those of Fig.~1b: 
(i) $s$--channel $\gamma,Z$ exchange
vertex diagrams involving loops built up by charged Higgs boson, chargino,
squark and slepton loops, $(ii)$ $t$--channel diagrams correcting the
$\Phi ee$ vertex involving chargino/sneutrino and neutralino/selectron
loops, and $(iii)$ box diagrams involving chargino/sneutrino and
neutralino/selectron states. \s 

For the associated production of the pseudoscalar state $A$, because of
CP invariance, there is no contribution from $W/Z$ bosons, charged Higgs
bosons and sfermions in the vertex diagrams and no contribution from
$W/Z$ bosons in the box diagrams. The process $\ee \ra A\gamma$ is
therefore mediated only by $s$--channel vertex diagrams involving
chargino and fermion loops, as well as $t$--channel vertex and box
diagrams involving chargino/sneutrino and neutralino/selectron states
(Fig.~1c). \s

The differential cross section for $\ee \ra \Phi \gamma$ with $\Phi=h,H$ 
and $A$ is given by the same formulae as in the SM, eq.~(2.1), with $M_H$ 
replaced by $M_\Phi$. Again, since by virtue of gauge invariance the form 
factor $C_3^\pm$ vanishes, the cross section with polarized $\ee$ beams
reduces as in eq.~(2.8) to
\beq
\frac{{\rm d}\;\sigma}{{\rm d}\;\cos\theta} &=& 
\frac{s-M_\Phi^2}{64 s\pi} \frac{1}{(16\pi^2)^2}  \bigg\{
 (1+\lambda_-)(1-\lambda_+) 
 \bigg[ u^2\; |C_1^+|^2 + t^2\; |C_2^+|^2 \bigg] \non\\
&& \mbox{}\hspace{2.5cm}
   + (1-\lambda_-)(1+\lambda_+) 
 \bigg[ u^2\; |C_1^-|^2 + t^2\; |C_2^-|^2 \bigg] \bigg\} \ . 
\eeq
In spite of this, we give $C_3^\pm$ to explicitly check gauge 
invariance. The form factors $C_{1,2}^\pm$ and $C_3^\pm$ are different 
for the production of CP--even and CP--odd Higgs production. We 
therefore discuss these two cases separately. 

\subsection*{3.1. CP--even Higgs boson production}

\subsubsection*{A. $s$--channel vertex corrections}

The one--loop $\gamma\gamma \Phi$ or $Z\gamma \Phi$ induced vertices can 
be written in the same tensorial form as in eq.~(2.10). For the CP--even
Higgs particles, the form factors $C_i^\pm$ are given in terms of the
amplitudes $G_{i}^\pm$ as in the SM by eqs.(2.11--2.12). In the MSSM, the 
various loop contributions to the form factors $G_{i}^\pm$ are given by 
\beq
G_{i}^{\gamma} & = & \frac{ e^3 M_W }{s_W} 
 \Bigg[ F_{i}^{\gamma,W} + F_{i}^{\gamma,H^+} 
 - \sum_{f}\, 4Q_f^2\,N_c\,\frac{m_f^2}{M_W^2} \,F^f_i  \Bigg]
 + \frac{ e^3 }{s_W}  \Bigg[ 
   F_{i}^{\gamma,\chi^+} 
  + \sum_{\tilde{f}} N_c Q_f F_{i}^{\gamma,\tilde{f}}  \Bigg]
\non \\
 G_{i}^{Z} & = & - \frac{ e^3 M_W}{c_W s_W^2} 
 \Bigg[ F_{i}^{Z,W}+F_{i}^{Z,H^+}
 - \sum_{f}\, 2Q_f\, N_c\,\frac{m_f^2}{M_W^2} 
(I^f_3 - 2 s_W^2 Q_f) \,F^f_i \Bigg]  \non \\
& & \vspace*{1cm}
- \frac{ e^3}{ s_W}  \Bigg[ 
   F_{i}^{Z,\chi^+} + \sum_{\tilde{f}} N_c Q_f F_{i}^{Z,\tilde{f}} \Bigg] , 
\eeq
where  $F_i^{V,W}$, $F_i^{V,H^+}$, $F_i^{V,\chi^+}$, $F_i^{V,\tilde{f}}$ 
and $F_i^{f}$  denote the $W$ boson, the charged Higgs boson, the chargino, 
the sfermion and the fermion contributions, respectively. 
Note that while the contributions of the fermions, charged Higgs bosons,
charginos and sfermions are separately gauge invariant, the 
contributions of the $W$ bosons are not [as in the SM] and the
contributions of the $t$--channel vertex and box diagrams are needed 
to insure the gauge invariance of this subset of diagrams. The contributions 
of the various  loops are as follows: 

\bigskip

\nn \underline{$W$ boson loops}: 

\bigskip

The contributions of the $W$ bosons to the $\gamma^* \gamma \Phi$ vertex
are
\beq
  F_{1}^{\gamma,W} &=& g_{\Phi G^+G^-} \frac{1}{c_W^2}\Big[ 
        M_W^2  C_0  - 4  C_{24} +  B_{13}   \Big]
+ g_{\Phi VV} \Big[
   M_{\Phi}^2    (  - 6 C_0 + C_{11} + C_{12} ) \non \\
&& + s (5 C_0 + C_{11} - C_{12}) + 24 C_{24} - B_{12}
       - 6 B_{13} + B_{23} \Big]\non \\
  F_{3}^{\gamma,W} &=& -
  \frac{1}{c_W^2} g_{\Phi G^+G^-} 4\; ( C_{12} + C_{23} )
   + g_{\Phi VV} 8\; ( 2 C_0 + 3( C_{12} + C_{23}) )
\eeq
and the contributions to the $Z^* \gamma \Phi$ vertex are 
\beq
 F_{1}^{Z,W} & = &  g_{\Phi VV} \Big[
  M_{\Phi}^2 c_W^2   (  - 6 C_0 + C_{11} + C_{12} + 2 C_{23} )
+ M_{\Phi}^2   (  - C_{11} + C_{23} ) \non \\
&&  + 2 M_W^2( 1- c_W^2 )   C_0 + c_W^2    (  - 1 + s (5 C_0 + C_{11} - C_{12} 
  + 2 C_{21} - 2 C_{23}) + 32 C_{24} \non \\
&& - B_{12} - 6 B_{13} - B_{23} )
 - 1/2 - s C_0 + s C_{11} + s C_{21} 
     - s C_{23} + B_{12} - B_{23} \Big] \non \\
&&  + g_{\Phi G^+G^-} \Big[ M_W^2 (1-c_W^{-2} ) C_0 
      + (1-c_W^{-2}/2 ) (   4 C_{24} - B_{13} ) \Big]
\non \\
 F_{3}^{Z,W} & = & g_{\Phi VV} \;4 \Big[
 ( 4 c_W^2 - 1 ) C_0 + ( 6 c_W^2 - 1 ) ( C_{12} +  C_{23} )  \Big] \non\\
&&  + g_{\Phi G^+G^-} \;2 \Big[ (c_W^{-2}-2) ( C_{12} + C_{23} ) \Big] \ .
\eeq
Note that the bosonic $(W, G^+)$ two--point function contribution to 
photon--$Z$ boson mixing, $F_{1}^{Z,W}|_{\rm mix}$ $=2 B_{23}\; g_{\Phi 
VV}$ is included. The two-- and three--point functions $B$ and $C$ are
the same as in eq.~(2.18) with $M_H \ra M_\Phi$. \s

At first sight, the couplings of the Higgs bosons to vector bosons and
charged Goldstones, which are given in Tab.~1a, look quite different.
However, in the MSSM the two couplings are intimately related; for
instance in the case of the light CP--even Higgs boson [a
similar identity holds for the heavy CP--even Higgs boson], one can
write\footnote{These expressions
are valid only at the tree--level, however the ratio should not be
altered by the radiative corrections in the Higgs sector if the latter
are consistently included in both the Higgs boson masses, the mixing
angle $\alpha$ and in the trilinear $g_{hG^+G^-}$ coupling.} 
these couplings in terms of the ratios $r_{h}=M_h^2/M_Z^2$ and 
$r_{H}=M_H^2/M_Z^2$ \cite{R1} 
\beq
g_{hVV} & \equiv & \sin (\beta-\alpha) = \left[ \frac{r_H (r_H 
-1)}{(r_H-r_h)(r_H+r_h-1)} \right]^{1/2} \non \\
g_{hG^+G^-} &\equiv & \cos2\beta \times \sin (\beta+\alpha) =
- \left[ \frac{ r_h r_H }{r_H+r_h-1} \right]^{1/2} \times
\left[ \frac{ r_h (r_H -1)}{r_H-r_h} \right]^{1/2} 
\eeq
and therefore 
\beq
\frac{ g_{hG^+G^-}}{g_{hVV}} = -\frac{M_h^2}{M_Z^2} \ .
\eeq
Inserting this identity, which is also valid in an arbitrary
two-Higgs Doublet Model (see for example \cite{S4a}),
in eqs.~(3.3--3.4), one then recovers the 
standard $W$ boson contributions eqs.~(2.16--2.17) with an overall 
additional factor $g_{\Phi VV}$.

\bigskip

\nn \underline{Fermion loops}: 

\bigskip

The contributions of the fermion loops are as in the SM, modulo a global 
factor $g_{\Phi ff}$ for the couplings of the Higgs bosons to fermions 
relative to the SM Higgs couplings (see Tab.~1b):
\beq
F^f_1 &=& \frac{g_{\Phi ff} }{2} 
 \Big[ 1 + (2 m_f^2 - M_{\Phi}^2 - s ) C_0 - 2 s(2 C_{11} + C_{21}) 
    + 2 ( s - M_{\Phi}^2) (C_{12}+ C_{23})  \Big]  \non \\
F^f_3 &=&   g_{\Phi ff} \Big[ C_0 + 4 C_{12} + 4 C_{23} \Big] \label{ffi}
\eeq
The three--point functions $C_0$, $C_{ij}$ are defined as in eq.~(2.15)
with $M_H \ra M_\Phi$. 

\bigskip

\nn \underline{Charged Higgs boson loops}: 

\bigskip

The contributions of the charged Higgs bosons to the $\gamma \gamma \Phi$ 
and $Z\gamma \Phi$ vertices are the same as those of the charged
Goldstones, and read
\beq
 F_{1}^{\gamma,H^+} &=& \left( \frac{1}{c_W^2} g_{\Phi H^+H^-}
      + 2 g_{\Phi VV} \right) ( 4 C_{24} -  B_{13} ) \non \\
 F_{3}^{\gamma,H^+} &=& \left( \frac{1}{c_W^2} g_{\Phi H^+H^-}
      + 2 g_{\Phi VV}\right) 4 ( C_{12} + C_{23} )
\eeq
\beq
 F_{1}^{Z,H^+} &=& \frac{1}{2} (2 c_W^2 - 1) \Big[ 2 g_{\Phi VV} 
 + \frac{1}{c_W^2} g_{\Phi H^+H^-} \Big] ( 4 C_{24} - B_{13} )
\non \\
F_{3}^{Z,H^+} &=& \frac{1}{2} (2 c_W^2 - 1) \Big[ 2 g_{\Phi VV} 
+ \frac{1}{c_W^2} g_{\Phi H^+H^-} \Big] \;4\;( C_{12} + C_{23} ) \ . 
\eeq
The couplings $g_{\Phi H^+ H^-}$ are the same as the charged Goldstone
couplings $g_{\Phi G^+ G^-}$ given in Tab.~1a at the tree level;
however, if the radiative corrections to the Higgs boson masses are
incorporated, one should also use the corrected coupling $g_{\Phi H^+
H^-}$ given (\ref{trileps}), to insure the gauge invariance of the result.
The two-- and three point functions are defined in this case as 
\beq
B_{13} &=& B_0(M_\Phi^2;M_{H^\pm}^2,M_{H^\pm}^2) \non\\
C_{ij} &=& C_{ij}(s,0,M_\Phi^2;M_{H^\pm}^2,M_{H^\pm}^2,M_{H^\pm}^2) \ . 
\eeq

\bigskip

\nn \underline{Chargino loops:} 

\bigskip

The chargino contributions $F_{i}^{V, \chi^+}$ with $i=1,3$ to the $V 
\gamma \Phi$ vertices can be cast into the compact form 
\beq
F_{i}^{V,\chi^+} & = & - 
\sum_{j,k=1,2} 
\left[ f_i \Big( -m_{\chi_j^+},-m_{\chi_k^+},-m_{\chi_k^+} \Big) 
 +  g_i \Big( -m_{\chi_k^+},-m_{\chi_j^+},-m_{\chi_j^+} \Big) \right]
\non\\ &&
\vspace*{1.4cm} \times \sum_{A,B=L,R} \, 
g_{V\chi^{+}_j\chi^{-}_k}^A  \;g_{\Phi\chi^{+}_k\chi^{-}_j}^B \ . 
\label{fichip}
\eeq
The couplings of the charginos to the Higgs bosons are given in 
Tab.~1c, while the chargino couplings to the photon and the $Z$ boson
read
\beq
g_{\gamma\chi^{+}_j\chi^{-}_k}^{L/R}=-\delta_{jk} \ \ {\rm and} \ \
g_{Z\chi^{+}_j\chi^{-}_k}^{L/R} = O^{\prime L/R} /(s_W c_W) \ . 
\eeq
The matrices $O^{\prime L/R}$ are defined as 
\beq
O^{\prime L}_{ij} = -V_{i1} V_{j1}^* -\frac{1}{2} V_{i2} V_{j2}^* 
+\delta_{ij} s_W^2
\ \ , \ \ 
O^{\prime R}_{ij} = -U_{i1} U_{j1}^* -\frac{1}{2} U_{i2} U_{j2}^* 
+\delta_{ij} s_W^2
\eeq
where the $U$ and $V$ matrices which diagonalize the chargino mass matrix 
can be found in Ref.~\cite{S5}. The functions $f_i/g_i \equiv 
f_i/g_i (m_1,m_2,m_3)$ are:
\beq
f_1 & = & m_1
   \Big[ 2 m_2 m_3 C_0 - M_{\Phi}^2 ( C_0 + C_{11} + 2 C_{12} + 2 C_{23})
      \non \\
&& \mbox{} \hspace{5mm}
   -  s( C_0 + 3  C_{11}
  - 2 C_{12} + 2 C_{21} - 2 C_{23} )  - 4 C_{24} + 1  \Big] \non \\
g_1 & = & m_2
   \Big[  M_{\Phi}^2 ( C_{11} + C_{12} + 2 C_{23} )
 + s ( C_{11} - C_{12}  + 2 C_{21} - 2 C_{23} ) 
 + 8 C_{24} - 1 \Big] \non  \\
&&  + m_3 \Big[ - M_{\Phi}^2( C_{12} + 2 C_{23} )
   - s (2 C_{11} - C_{12} + 2 C_{21}
     - 2 C_{23} ) - 4 C_{24} + 1 \Big]
\non \\
f_3 & = &  2 m_1
         \Big[  C_0 +  C_{11} + 2 C_{12} + 2 C_{23} \Big]
\non \\
g_3 & = & 2
      \Big[  - m_2 ( C_{11} - C_{12} ) + m_3 ( C_{12} + 2 C_{23} ) \Big]
\eeq
The arguments of the three--point functions $C_0$ and $C_{ij}$ are 
specified according to 
\beq
C_{ij} = C_{ij}(s,0,M_\Phi^2;m_{\chi^+}^2,m_{\chi^+}^2,m_{\chi^+}^2) \ . 
\eeq

\bigskip 

\nn \underline{Sfermion loops:} 

\bigskip

Finally, the squark/slepton contribution to the $V \gamma \Phi$ vertices
can be written as 
\beq
F_i^{V,\tilde{f} } & = & 
\sum_{j,k=1,2} \;
g_{\Phi\tilde{f}_j\tilde{f}_k} \; g_{V\tilde{f}_k\tilde{f}_j} \; s_i
\eeq
This term comes only from Higgs emission from the sfermion lines
since the contributions of the two-- and one--point $Z$--$\gamma$ 
mixing diagrams cancel each other. The functions $s_{i} \equiv 
s_{i} (m_{\tilde{f}_j},m_{\tilde{f}_k},m_{\tilde{f}_k})$ read:
\beq
s_1 &=& 2B_{0}(M_{\Phi}^2; m_{\tilde{f}_k}^2, m_{\tilde{f}_k}^2)
-8C_{24}(s,0, M_\Phi^2;m_{\tilde{f}_j}^2, m_{\tilde{f}_k}^2,m_{\tilde{f}_k}^2)
\non \\
s_3 &=& -8[
 C_{12}(s,0,M_\Phi^2;m_{\tilde{f}_j}^2,m_{\tilde{f}_k}^2,m_{\tilde{f}_k}^2)
+C_{23}(s,0,M_\Phi^2;m_{\tilde{f}_j}^2,m_{\tilde{f}_k}^2,m_{\tilde{f}_k}^2) ] 
\ . 
\eeq
The squark couplings to the Higgs bosons, including mixing \cite{S6} between
left-- and right--handed sfermions\footnote{This mixing is proportional to the
fermion mass and in practice, is non negligible only for the partners of
the third generation fermions.} are 
\beq
g_{\Phi \tilde{f}_1\tilde{f}_1} & = & 
  C_{LL}^\Phi \cos^2\theta_f + C_{RR}^\Phi \sin^2\theta_f 
   + 2 C_{RL}^\Phi \cos\theta_f\sin\theta_f \non \\ 
g_{\Phi \tilde{f}_2\tilde{f}_2} & = & 
   C_{RR}^\Phi \cos^2\theta_f + C_{LL}^\Phi \sin^2\theta_f
   - 2 C_{RL}^\Phi \cos\theta_f\sin\theta_f \non \\
g_{\Phi \tilde{f}_1\tilde{f}_2} & = & 
  C_{RL}^\Phi (\cos^2\theta_f - \sin^2\theta_f)
   + (C_{RR}^\Phi -C_{LL}^\Phi ) \cos\theta_f \sin\theta_f 
\eeq
with
\beq
 C_{LL}^\Phi & = & - \frac{M_Z}{c_W}(I_3^f-Q_f s_W^2) g_2^\Phi 
    - \frac{m_f^2}{M_W} g_1^\Phi 
\non \\
 C_{RR}^\Phi & = & - \frac{M_Z}{c_W} (Q_f s_W^2) g_2^\Phi 
    - \frac{m_f^2}{M_W}  g_1^\Phi
\non \\
 C_{RL}^\Phi & = & - \frac{m_f}{2 M_W} 
 \left[ A_f g_4^\Phi + \mu g_3^\Phi \right] \ , 
\eeq
where $A_f$ and $\mu$ are the soft--SUSY breaking trilinear term
and the Higgs--higgsino mass parameter, respectively. The factors
$g_i^\Phi$ are given in Tab.~2a, and in the limit of zero--mixing
the couplings reduce to those given in Tab.~2b. The photon coupling 
to sfermions is just $ g_{\gamma \tilde{f}_i \tilde{f}_j} = 
\delta_{ij} Q_f$, while the couplings of the $Z$ boson to sfermions in 
the case of mixing are given by 
\beq
g_{Z \tilde{f}_1 \tilde{f}_1}  & = & 
D_{LL}\cos^2\theta_f + D_{RR}\sin^2\theta_f \non\\
g_{Z \tilde{f}_2 \tilde{f}_2}  & = & 
D_{RR}\cos^2\theta_f + D_{LL}\sin^2\theta_f \non\\
g_{Z \tilde{f}_1 \tilde{f}_2}  & = & 
  (D_{RR}-D_{LL}) \sin\theta_f\cos\theta_f 
\eeq
with 
\beq
D_{LL} = \frac{1}{s_W c_W} (I^f_3-Q_f s_W^2) \ \ \ {\rm and} \ \
D_{RR} = \frac{1}{s_W c_W}(-Q_f s_W^2) \ . 
\eeq

\subsubsection*{B. $t$--channel vertex corrections}

In addition to the contribution of the $W/\nu$ and $Z/e$ SM like loops,
there are two additional contributions to the $t$--channel $\Phi ee$
vertex diagrams: one with chargino/sneutrino loops and another one with
neutralino/selektron loops. In the case where the mixing in the
selectron sector is neglected, all these diagrams do not contribute to
the form factors $C_{1,2}^{e \pm}$:
\beq
C_{1,2}^{e\pm}=0 \ . 
\eeq
There is, however, a contribution to the form factors $C_3^{\pm}$ which 
reads 
\beq
C_3^{e+} &=& \frac{e^4}{s_W^3} \Bigg[ 
\frac{M_W}{2}\;g_{\Phi VV}\; f_3^{fVV}(M_W,0,M_W)
-\frac{M_Z}{4c_W^3}\;g_{\Phi VV}\;(z^+)^2 f_3^{fVV}(M_Z,m_e,M_Z)
\non\\ &&\mbox{}\hspace{5mm}
+\sum_{A=R,L} \sum_{j,k}\; g_{e\chi_j^+\nu}\; g_{e\chi_k^+\nu}
  g_{\Phi\chi_j^+\chi_k^-}^A\; 
  f_{3A}^{ffS}(-m_{\chi_j^+},m_{\tilde{\nu}},-m_{\chi_k^+})
\non\\ &&\mbox{}\hspace{5mm}
- \frac{M_Z}{2c_W}\sum_{j}\; g_{e\chi_j^+\nu}\; g_{e\chi_j^+\nu}
  g_{\Phi\tilde{\nu}\tilde{\nu}}\; 
  f_3^{fSS}(m_{\tilde{\nu}},-m_{\chi_j^+},m_{\tilde{\nu}})
\non\\ &&\mbox{}\hspace{5mm}
+\frac{1}{2}\sum_{A=R,L} \sum_{j,k}\;
  g_{e\chi_j^0\tilde{e}_L}\; g_{e\chi_k^0\tilde{e}_L}\;
  g_{\Phi\chi_j^0\chi_k^0}^A\; 
  f_{3A}^{ffS}(m_{\chi_j^0},m_{\tilde{e}_L},m_{\chi_k^0})
\non\\ &&\mbox{}\hspace{5mm}
+ \frac{M_Z}{4c_W}\sum_{j}\;
  g_{e\chi_j^0\tilde{e}_L}\; g_{e\chi_j^0\tilde{e}_L}\;
  g_{\Phi\tilde{e}_L\tilde{e}_L}\; 
  f_3^{fSS}(m_{\tilde{e}_L},m_{\chi_j^0},m_{\tilde{e}_L})
\Bigg] + \mbox{\rm crossed}
\eeq
\beq
C_3^{e-} &=& \frac{e^4}{s_W^3} \Bigg[ 
-\frac{M_Z}{4c_W^3}\;g_{\Phi VV}\;(z^-)^2 f_3^{fVV}(M_Z,m_e,M_Z)
\non\\ &&\mbox{}\hspace{5mm}
+\frac{1}{2}\sum_{A=R,L} \sum_{j,k}\;
  g_{e\chi_j^0\tilde{e}_R}\; g_{e\chi_k^0\tilde{e}_R}\;
  g_{\Phi\chi_j^0\chi_k^0}^A\; 
  f_{3\bar{A}}^{ffS}(m_{\chi_j^0},m_{\tilde{e}_R},m_{\chi_k^0})
\non\\ &&\mbox{}\hspace{5mm}
+ \frac{M_Z}{4c_W}\sum_{j}\;
  g_{e\chi_j^0\tilde{e}_R}\; g_{e\chi_j^0\tilde{e}_r}\;
  g_{\Phi\tilde{e}_R\tilde{e}_R}\; 
  f_3^{fSS}(m_{\tilde{e}_R},m_{\chi_j^0},m_{\tilde{e}_R})
\Bigg] + \mbox{\rm crossed}
\eeq
with $\bar{A} = L,R$ when $A = R,L$ and with the functions $f_{3}^{...}
\equiv f_3^{...} (m_1, m_2, m_3)$ given by 
\beq
f_3^{fSS} &=& -\frac{1}{2} C_{12}  \non\\
f_3^{fVV} &=&  C_{12}   \non\\
f_{3L}^{ffS} &=& \frac{1}{2} m_3 C_{12} \non \\
f_{3R}^{ffS} &=& \frac{1}{2} m_1 ( C_0 + C_{12} )
\eeq
involving the three--point functions $C_{ij}$ specified as 
\beq
C_{ij}= C_{ij}(m_e^2, t, M_\Phi^2; m_1^2, m_2^2, m_3^2) 
\eeq
for the direct diagrams; for the crossed diagrams one has to make the
substitution $t \ra u$ in the previous equation. The couplings of 
the Higgs bosons to vector bosons, charginos and neutralinos 
are given in Tab.~1, and those to sleptons in Tab.~2b. The only
remaining couplings to be defined are the 
electron--chargino--sneutrino and electron--neutralino--selectron
couplings; normalized to $g_{W}=\left[\sqrt{2}G_F\right]^{ 1/2}$
$M_W$, they are given by
\beq
g_{e \chi_i^+ \tilde{\nu}} = V_{i1} \ \ , \ \
g_{e \chi_i^0 \tilde{e}_R} = 2\frac{s_W}{c_W} N_{i2} \ \ , \ \
g_{e \chi_i^0 \tilde{e}_L} = -N_{i2} -\frac{s_W}{c_W} N_{i1} \ .
\eeq
The matrices $V$ and $N$ which diagonalize the chargino and
neutralino mass matrices can be found in Ref.~\cite{S5}.

\subsubsection*{C. Box corrections}

The box diagrams involve the contributions form $W$/neutrino and $Z$/electron
loops as in the case of the SM Higgs boson, plus chargino/sneutrino and 
neutralino/selectron loops. The contributions to $C_i^{\rm box \pm}$ read:
\beq
C_i^{\rm box+} &=& \frac{e^4}{s_W^3} \Bigg[ 
 -\frac{M_W}{2} \,f_i^{fVVV}(M_W,0,M_W,M_W) -\frac{M_W}{2} 
 f_i^{fVVS}(M_W,0,M_W,M_W) ) 
\non\\ && \mbox{}\hspace{6mm} + \sum_j\; 
\frac{M_Z}{4c_W} g_{e\chi^0_j\tilde{e}_L}\; g_{e\chi^0_j\tilde{e}_L}\;
g_{\Phi\tilde{e}_L\tilde{e}_L}\;f_i^{fSSS}
 (m_{\tilde{e}_L},m_{\chi^0_j},m_{\tilde{e}_L},m_{\tilde{e}_L}) \non\\
&&\mbox{}\hspace{6mm} 
 - \sum_{A=R,L} \sum_{j,k} \; g_{e\chi^+_j\tilde{\nu}} \;
  g_{e\chi^+_k\tilde{\nu}} \; g_{\Phi\chi^+_j\chi^-_k}^A \;
  f_{iA}^{fffS}(-m_{\chi^+_j},m_{\tilde{\nu}},-m_{\chi^+_k},-m_{\chi^+_j}) 
\Bigg]
+\mbox{\rm cros.} \non\\
&& + \frac{e^4}{s_W^3} \Bigg[ 
 - \frac{ M_Z }{4\,c_W^3} (z^+)\,f_i^{ffVV}(m_e,M_Z,M_Z,m_e) \non\\
&& \mbox{}\hspace{1cm} + \sum_{A=R,L} \sum_{j,k} \frac{1}{2}
  g_{e\chi^0_j\tilde{e}_L}\; g_{e\chi^0_k\tilde{e}_L}\;
  g_{\Phi\chi^0_j\chi^0_k}^A\;
  g_{iA}^{ffSS}(m_{\chi^0_j},m_{\tilde{e}_L},m_{\tilde{e}_L},m_{\chi^0_k})
\non\\ &&\mbox{}\hspace{1cm}  + \sum_{j,k} \frac{M_Z}{2c_W}
  g_{e\chi^+_j\tilde{\nu}}\; g_{e\chi^+_k\tilde{\nu}}\;
  g_{\Phi\tilde{\nu}\tilde{\nu}}\;
  f_i^{ffSS}(-m_{\chi^+_j},m_{\tilde{\nu}},m_{\tilde{\nu}},-m_{\chi^+_k})
\Bigg]
\non
\\[2mm]
C_i^{\rm box-} &=& \frac{e^4}{s_W^3} \Bigg[ 
\sum_j\; \frac{M_Z}{4c_W} g_{e\chi^0_j\tilde{e}_R}\; 
g_{e\chi^0_j\tilde{e}_R}\;
g_{\Phi\tilde{e}_R\tilde{e}_R}\;f_i^{fSSS} 
(m_{\tilde{e}_R},m_{\chi^0_j},m_{\tilde{e}_R},m_{\tilde{e}_R}) \Bigg]
+\mbox{\rm cros.} \non\\
&& + \frac{e^4}{s_W^3} \Bigg[ 
 - \frac{ M_Z }{4\,c_W^3} (z^-)\,f_i^{ffVV}(m_e,M_Z,M_Z,m_e) \non\\
&& \mbox{}\hspace{1cm} + \sum_{A=R,L} \sum_{j,k}
  g_{e\chi^0_j\tilde{e}_R}\; g_{e\chi^0_j\tilde{e}_R}\;
  g_{\Phi\chi^0_j\chi^0_k}^A\;
 g_{i\bar{A}}^{ffSS}(m_{\chi^0_j},m_{\tilde{e}_R},m_{\tilde{e}_R},m_{\chi^0_k})
\Bigg]
\eeq
where the sums run over the two chargino and the four neutralino states.
All the couplings have been previously defined, and the box functions 
$f_i^{fxyz}=f_i^{fxyz}(m_1^2,m_2^2,m_3^2,m_4^2)$ are given by
\beq
f_1^{fSSS} &=&    - D_{23} + D_{25}  \non\\
f_2^{fSSS} &=&    - D_{23} + D_{26}  \non\\
f_3^{fSSS} &=& \frac{1}{2} 
  [ t ( D_{23} -  D_{26} ) + u ( D_{23} -  D_{25} ) + 2 D_{27} ] 
 \\[2mm]
f_1^{fVVS} &=& 2 ( D_0 + D_{11} ) \non\\
f_2^{fVVS} &=& 0  \non\\
f_3^{fVVS} &=& \frac{1}{2} 
[  - 2 u ( D_0 + D_{11} ) - ( s + 2 t ) D_{12} + ( s + 2 t + u ) D_{13}
 + ( s + t + u ) D_{23} \non\\
&& \mbox{}\hspace{1cm} + s D_{24} - (s+u) D_{25} - (s+t) D_{26} )
  + 4 D_{27} ] 
\\[2mm]
f_1^{fVVV} &=& D_0 + D_{11} - 2 D_{13} + 2 D_{23} - 2  D_{25} 
\non \\
f_2^{fVVV} &=& 2 [  - D_{12} + D_{13} + D_{23} - D_{26} ]
\non \\
f_3^{fVVV} &=& \frac{1}{2} 
 [  u ( D_{0} + D_{11} ) - ( s - t ) D_{12} + ( s - t + u ) D_{13} 
  - 2 (  s + 2 t + 2 u ) D_{23} \non\\
&&  \mbox{}\hspace{1cm}- 2 s D_{24} + 2( s + 2 u ) D_{25} 
  + 2( s + 2 t ) D_{26} - 12 D_{27} ]
\\[2mm]
f_{1L}^{fffS} &=&  - m_1 ( D_0 + D_{11} ) + m_4 ( D_{13}
           - D_{23} + D_{25} ) \non\\
f_{1R}^{fffS} &=&  m_3 (  - D_{23} + D_{25} )\non \\
f_{2L}^{fffS} &=&  m_4 (  - D_{23} + D_{26} )\non\\
f_{2R}^{fffS} &=&  m_3 ( D_{12} - D_{13} - D_{23} + D_{26} ) \non \\
f_{3L}^{fffS} &=&  \frac{1}{2}
 [  ( m_1 - m_4 ) s D_{12} - ( m_1 - m_4 ) s D_{13}
 + ( m_1 ( s + u + t ) - m_4 s ) D_{23}  \non\\
&& \mbox{}\hspace{1cm} + ( m_1 - m_4 ) s D_{24}
 - ( ( m_1 - m_4 ) s + m_1 u ) D_{25}  \non\\
&& \mbox{}\hspace{1cm} - ( ( m_1 - m_4 ) s + m_1 t ) D_{26}
 + ( 4 m_1 - 2 m_4 ) D_{27}  ] \non\\
f_{3R}^{fffs} &=&   \frac{1}{2} m_3
  (  m_1 m_4 D_0 - s ( D_{23} + D_{24} - D_{25} - D_{26} ) - 2 D_{27} )
\label{ffffs}
\eeq
with the four--point functions 
\beq
D_{ij} = D_{ij}(m_e^2,m_e^2,M_{\Phi}^2,0,s,u;m_1^2,m_2^2,m_3^2,m_4^2) \ .
\eeq
The remaining box functions read
\beq
f_1^{ffVV} &=&  2 ( - D_{11} + D_{12} + D_{22} - D_{24} ) 
\non\\
f_2^{ffVV} &=& 2 ( D_{22} - D_{26} )\non \\
f_3^{ffVV} &=&   
- m_1 m_4 D_0 + s ( D_{22} - D_{24} + D_{25} - D_{26} ) + 2 D_{27} 
\\[2mm]
f_1^{ffSS } &=&  - D_{22} + D_{24} \non \\
f_2^{ffSS } &=&  - D_{12} + D_{13} - D_{22} + D_{26} \non \\
f_3^{ffSS } &=&  \frac{1}{2} [ m_1 m_4 D_0 - s ( D_{22}
         - D_{24} + D_{25} - D_{26} ) - 2 D_{27} ]
\eeq
with the four--point functions 
\beq
D_{ij}=D_{ij}(m_e^2,M_{\Phi}^2,m_e^2,0,t,u;m_1^2,m_2^2,m_3^2,m_4^2) \ ,
\eeq
and
\beq
g_{1L}^{ffSS} &=&  m_1 (  D_0 +  D_{11} +  D_{12} + D_{24} ) \non\\
g_{1R}^{ffSS} &=&  m_4 (  D_{12} +  D_{24} )   \non \\
g_{2L}^{ffSS} &=&  m_1 (  D_{13} +  D_{26} )  \non \\
g_{2R}^{ffSS} &=&  m_4 ( D_{26} )   \non \\
g_{3L}^{ffSS} &=&  \frac{1}{2} m_1
 (  - u ( D_0 + D_{11} + D_{12} + D_{24} )
     - t ( D_{13} + D_{26} ) - 2 D_{27} ) \non\\
g_{3R}^{ffSS} &=&  \frac{1}{2} m_4
   ( -  u ( D_{12} + D_{24} ) -  t D_{26} - 2  D_{27} )
\label{gffss}
\eeq
with
\beq
D_{ij}=D_{ij}(m_e^2,0,m_e^2,M_{\Phi}^2,u,t;m_1^2,m_2^2,m_3^2,m_4^2) \ .
\eeq
The contribution of the crossed diagrams can be obtained by 
simply interchanging $t \leftrightarrow u$ in the relevant 
expressions given above. \s

The gauge invariance of the complete result has been checked
explicitly. The sum of $W$/neutrino and $Z$/electron contributions to
the $t$-channel vertex and box diagrams and the contributions of the
$W$ bosons to the $s$-channel vertices are gauge invariant as in the
SM. For the neutralino/selectron and the chargino/sneutrino
contributions, only the sum $C^e +C^{\rm box}$ is gauge invariant. 

\subsection*{3.2. CP--odd Higgs production}

In the case of the CP--odd Higgs bosons, fewer diagrams contribute
to the associated production with a photon compared to the case of the
CP--even Higgs particles. Indeed, because of CP invariance, there is no
pseudoscalar couplings to vector bosons, to charged Higgs bosons and to the
SUSY partners of the light fermions. The mixing between left--handed and
right--handed third generation sfermions will induce a $A \tilde{f}_1
\tilde{f}_2$ coupling, however, CP--invariance still imposes the sum of
all sfermion loops to vanish. 

\subsubsection*{A. $s$--channel vertex corrections}

The one--loop $\gamma\gamma A$ or $Z\gamma A$ vertices can be written 
in the same tensorial form as in eq.~(2.10). However, for the CP--odd Higgs 
boson, the form factors $C_{1,2}^\pm$ are different because now the
amplitude $G_6$ gives a non--vanishing contribution. Using the same 
notation as previously, we have for the $Z\gamma A$ vertex
\beq
C^{Z\;\pm}_1 &=&
 \frac{e\, z^{\pm}}{4c_W s_W} \, \frac{1}{s-M_Z^2}\; (- G_3^{Z} \pm 
G_6^{Z})\non\\ 
C^{Z\;\pm}_2 &=&
 \frac{e\, z^{\pm}}{4c_W s_W}\, \frac{1}{s-M_Z^2}\; (- G_3^{Z} \mp 
G_6^{Z})\non\\ 
C^{Z\;\pm}_3 &=&
  \frac{e\, z^{\pm}}{4c_W s_W}\,\frac{1}{s-M_Z^2} \; 
 \left( G_1^{Z} - \frac{s-M_H^2}{2}\;G_3^{Z}\right) \ , 
\eeq
and for the $\gamma\gamma A$ vertex
\beq
C^{\gamma\;\pm}_1 &=&
  \frac{e}{2}\, \frac{1}{s}\; ( - G_3^{\gamma} \pm G_6^{\gamma}) \non \\ 
C^{\gamma\;\pm}_2 &=&
  \frac{e}{2}\, \frac{1}{s}\; ( - G_3^{\gamma} \mp G_6^{\gamma}) \non \\ 
C^{\gamma\;\pm}_3 &=&
  \frac{e}{2}\, \frac{1}{s}\; 
\left( G_1^{\gamma} - \frac{s-M_H^2}{2}\;G_3^{\gamma}\right) \ .
\eeq

Only heavy fermion and chargino loops contribute to the form factors 
$G_{i}^{\gamma}$ and $G_{i}^{Z}$. They are separately gauge invariant 
and are decomposed according to
\beq
 G_{i}^{\gamma} & = & - \frac{ e^3 M_W }{s_W}  \;
  \sum_{f}\, 4Q_f^2\,N_c\,\frac{m_f^2}{M_W^2} \,F^f_i
+\frac{ e^3 }{s_W}  \; F_{i}^{\gamma,\chi^+}  
\non\\
 G_{i}^{Z} & = & 
   \frac{ e^3 M_W}{c_W s_W^2} \;
   \sum_{f}\, 2Q_f\, N_c\,\frac{m_f^2}{M_W^2} 
(I^f_3 - 2 s_W^2 Q_f) \,F^f_i        \Bigg] 
  - \frac{e^3}{s_W} \; F_{i}^{Z,\chi^+}  \ . 
\eeq

The fermion loop contribution to the $V A \gamma$ vertex $F_i^f$ is 
given by 
\beq
F^f_1 &=&  F_{3}^f = 0 \non \\
F^f_6 &=& - g_{A ff} C_{0} (s,0,M_A^2;m_f^2,m_f^2,m_f^2)
\eeq
with the couplings $g_{Aff}$ in Tab.~1a. The chargino loops
yield 
\beq
F_{1,3}^{V,\chi^+} & = & -  \sum_{j,k=1,2} 
\left[ f_{1,3} \Big( -m_{\chi_j^+},-m_{\chi_k^+},-m_{\chi_k^+} \Big) 
 + g_{1,3} \Big( -m_{\chi_k^+},-m_{\chi_j^+},-m_{\chi_j^+} \Big) \right]
\non\\ && \hspace*{1.4cm} \times 
 \Big( g_{V\chi^{+}_j\chi^{-}_k}^R + g_{V\chi^{+}_j\chi^{-}_k}^L \Big) 
 \Big( g_{A\chi^{+}_k\chi^{-}_j}^R + g_{A\chi^{+}_k\chi^{-}_j}^L \Big) 
\non \\
F_{6}^{V,\chi^+} &=&   
\sum_{j,k=1,2} 
 \left[ f_6 \Big( -m_{\chi_j^+},-m_{\chi_k^+},-m_{\chi_k^+} \Big) 
 -  g_6 \Big( -m_{\chi_k^+},-m_{\chi_j^+},-m_{\chi_j^+} \Big) \right]
\non\\ && \hspace*{1.4cm} \times
 \Big( g_{V\chi^{+}_j\chi^{-}_k}^R + g_{V\chi^{+}_j\chi^{-}_k}^L \Big) 
 \Big( g_{A\chi^{+}_k\chi^{-}_j}^R - g_{A\chi^{+}_k\chi^{-}_j}^L \Big) 
%\left( - g_{V\chi^{+}_j\chi^{-}_k}^L \; g_{A\chi^{+}_k\chi^{-}_j}^R
%       + g_{V\chi^{+}_j\chi^{-}_k}^R \; g_{A\chi^{+}_k\chi^{-}_j}^L
%       + g_{V\chi^{+}_j\chi^{-}_k}^L \; g_{A\chi^{+}_k\chi^{-}_j}^L
%       - g_{V\chi^{+}_j\chi^{-}_k}^R \; g_{A\chi^{+}_k\chi^{-}_j}^R  \right)
\non \\
\eeq
with $f_{1,3}$ and $g_{1,3}$ given by eqs.~(3--14) and the new functions 
\beq
f_6 &=& 2 m_1 ( C_0 + C_{11} ) \non\\
g_6 &=& 2 ( m_2 C_{11} - m_2 C_{12} + m_3 C_{12} ) \ .
\eeq
The arguments of the $C_{ij}$ functions are specified as in eq.~(3.15) 
with $M_\Phi \equiv M_A$.

\subsubsection*{B. $t$--channel vertex corrections}
\bigskip

The $t$--channel $Aee$ vertex corrections are built--up only by
chargino/sneutrino and neutralino/selectron loops since there is
no $AVV$ coupling. Again, these loops do not contribute to $C_{1,2}^e$ 
and the expression of the contribution to $C_3^e$ is simpler than
in the case of the CP--even Higgs bosons, since in the absence of
slepton mixing, there is no $A\tilde{l} \tilde{l}$ coupling:
\beq
C_{1,2}^{e\pm}&=&0\\
C_3^{e+} &=& \frac{e^4}{s_W^3} \Bigg[ 
\sum_{a=R,L} \sum_{j,k}\; g_{e\chi_j^+\nu}\; g_{e\chi_k^+\nu}
  g_{A\chi_j^+\chi_k^-}^a\; 
f_{3a}^{ffS}(-m_{\chi_j^+},m_{\tilde{\nu}},-m_{\chi_k^+})
\non\\ && \mbox{}\hspace{5mm}
+\frac{1}{2}\sum_{a=R,L} \sum_{j,k}\;
  g_{e\chi_j^0\tilde{e}_L}\; g_{e\chi_k^0\tilde{e}_L}\;
  g_{A\chi_j^0\chi_k^0}^a\; 
  f_{3a}^{ffS}(m_{\chi_j^0},m_{\tilde{e}_L},m_{\chi_k^0})
\Bigg] + \mbox{\rm crossed}
\non\\
C_3^{e-} &=& \frac{e^4}{s_W^3} \Bigg[ 
\frac{1}{2}\sum_{a=R,L} \sum_{j,k}\;
  g_{e\chi_j^0\tilde{e}_R}\; g_{e\chi_k^0\tilde{e}_R}\;
  g_{A\chi_j^0\chi_k^0}^a\; 
  f_{3\bar{a}}^{ffS}(m_{\chi_j^0},m_{\tilde{e}_R},m_{\chi_k^0})
\Bigg] + \mbox{\rm crossed} \non \\
\eeq
with the functions $f_{3a}^{fSS}$ [$\bar{a} = L$ when $a = R$ and 
vice versa] are given by eq.~(3.25) with the same $C_{ij}$ functions
as in eq.~(3.26) with the replacement $M_\Phi \ra M_A$, and the crossed
contributions are obtained by interchanging $u$ and $t$. 

\subsubsection*{C. Box corrections}

Only the chargino/sneutrino and neutralino/selectron boxes contribute to
the production of the pseudoscalar. Using the same notation as in the case 
of the CP--even Higgs bosons, the $C_i^{\rm box \pm}$ amplitudes are given by
\beq
C_i^{\rm box+} &=& \frac{e^4}{s_W^3} \Bigg[ - \sum_{a=R,L} \sum_{j,k} \; 
  g_{e\chi^+_j\tilde{\nu}}\; g_{e\chi^+_k\tilde{\nu}}\;
  g_{A\chi^+_j\chi^-_k}^a\;
  f_{ia}^{fffS}(-m_{\chi^+_j},m_{\tilde{\nu}},-m_{\chi^+_k},-m_{\chi^+_j}) 
\Bigg] +\mbox{\rm cr.} \non\\
&& + \frac{e^4}{s_W^3} \frac{1}{2} \sum_{a=R,L} \sum_{j,k}
  g_{e\chi^0_j\tilde{e}_L}\; g_{e\chi^0_k\tilde{e}_L}\;
  g_{A\chi^0_j\chi^0_k}^a\;
 g_{ia}^{ffSS}(m_{\chi^0_j},m_{\tilde{e}_L},m_{\tilde{e}_L},m_{\chi^0_k})
\non \\ 
C_i^{\rm box-} &=& \frac{e^4}{s_W^3} \frac{1}{2} \sum_{a=R,L} \sum_{j,k}
  g_{e\chi^0_j\tilde{e}_R}\; g_{e\chi^0_k\tilde{e}_R}\;
  g_{A\chi^0_j\chi^0_k}^a\;
 g_{i\bar{a}}^{ffSS}(m_{\chi^0_j},m_{\tilde{e}_R},m_{\tilde{e}_R},m_{\chi^0_k})
\eeq
where sums run over the two chargino and the four neutralino states
[again $\bar{a}=L,R$ when $a=R,L$]. The various couplings and the box
functions $f_i^{....}$ and $g_i^{....}$ have been given previously
eqs.~(\ref{ffffs}--\ref{gffss}); the crossed contribution is obtained 
by the interchange $u \leftrightarrow t$. \s

Again, we have checked explicitely that the sum of the contributions 
$C^e + C^{\rm box}$ is indeed gauge invariant. 

\subsection*{3.3. Results}

Before discussing our numerical results, let us first shortly describe
our parametrization of the MSSM parameter space. Besides the four
masses, $M_h, M_H, M_A$ and $M_{H^\pm}$, the Higgs sector is described
at the tree level by two additional parameters, $\tb$ and a mixing
angle $\alpha$. Due to supersymmetry constraints only two of them are 
independent and the two inputs are in general taken to be $\tb$ and $M_A$. 
Radiative corrections \cite{R15,R15b}, the leading part of which grow as 
the fourth power of the top mass and logarithmically with the common squark 
mass, change significantly the relations between masses and couplings
and shift the mass of the lightest CP--even Higgs boson upwards. These 
radiative corrections are very important and should therefore be included 
in the analysis. We will, however, only include the leading part of 
this correction which in the simplest case can be pa\-ram\-e\-trized in 
terms of the quantity \cite{R15} 
\beq
\epsilon = \frac{3 G_{F}}{\sqrt{2}\pi^2} \frac{m_t^4}{\sin^2\beta}
\log\left( 1+\frac{M_{\tilde{q}}^2}{m_t^2} \right) \ . 
\eeq
The CP--even Higgs boson masses are then given in terms of the 
pseudoscalar mass $M_A$ and $\tb$ as
\beq
M_{h,H}^2 &=&\frac{1}{2}\Bigg[ M_A^2+M_Z^2+\epsilon \non\\
&& \mbox{} \mp \sqrt{(M_A^2+M_Z^2+\epsilon)^2- 4M_A^2 M_Z^2 \cos^2 2\beta 
   - 4\epsilon( M_A^2 \sin^2 \beta + M_Z^2 \cos^2 \beta)} 
   \Bigg] \non \\
\eeq
and the mixing angle $\alpha$ by
\beq
\tan 2\alpha &=& \tan 2\beta 
 \frac{M_A^2 + M_Z^2}{M_A^2 - M_Z^2 +\epsilon/\cos 2\beta} \ , \hspace*{1cm}
- \frac{\pi}{2} \leq \alpha \leq 0 \ . 
\eeq
Once $\tb$ and $M_A$ are chosen and the leading radiative correction is
included in $\alpha$ and $M_{h}/M_H$, all the couplings of the Higgs
bosons to fermions, gauge bosons, charginos, neutralinos and sfermions
are fixed; these couplings are given in Tables 1 and 2. There is,
however, one exception: the trilinear Higgs couplings also receive
radiative corrections which are not completely mapped into the
corrections to the angle $\alpha$. The only trilinear couplings which we
will need in our analysis are the CP--even Higgs boson couplings to
charged Higgs bosons and Goldstones [which are the same at the Born level]; 
to leading order they are given by \cite{S7}
\beq
g_{h H^+H^-} &=& \cos 2\beta \sin(\beta+\alpha)
 + \frac{\epsilon}{M_Z^2} \frac{\cos\alpha \cos^2\beta}{\sin\beta} \non\\
g_{H H^+H^-} &=& -\cos 2\beta \cos(\beta+\alpha)
 + \frac{\epsilon}{M_Z^2} \frac{\sin\alpha \cos^2\beta}{\sin\beta} \non\\[2mm]
g_{h G^+G^-} &=& \cos 2\beta \sin(\beta+\alpha)
 + \frac{\epsilon}{M_Z^2} \cos\alpha \sin\beta \non\\
g_{H G^+G^-} &=& -\cos 2\beta \cos(\beta+\alpha)
 + \frac{\epsilon}{M_Z^2} \sin\alpha \sin \beta
\label{trileps}
\eeq
It is these expressions which have to be used in order to obtain the 
same results for the lightest CP--even Higgs boson $h$ and for the SM 
Higgs boson $H^0$ in the decoupling limit, and to insure the gauge 
invariance of the final results. \s

In the MSSM, there are two charginos $\chi_i^+ [i=1,2$] and four
neutralinos $\chi_{i}^0$ [$i=1$--4]. Their masses and their couplings
are given in terms of the Higgs--higgsino mass parameter $\mu$, the
ratio of the vacuum expectation values $\tb$ and the wino mass parameter
$M_2$. The bino and gluino masses are related to $M_2$ by
$M_1 \sim M_2/2$ and $m_{\tilde{g}} \sim 3.5 M_2$, when the gaugino
masses and the three coupling constants of
SU(3)$\times$SU(2)$\times$U(1) are unified at the GUT scale. The squark
and slepton masses are given in terms of the parameters $\mu$ and $\tb$,
as well as the left-- and right--handed scalar masses $M_{\tilde{f}_L}$
and $M_{\tilde{f}_R}$, which we will take to be equal, and the
soft--SUSY breaking trilinear coupling $A_f$. In the case of the third
generation sfermions, mixing between left-- and right--handed states is 
taken into account; in pratice it is important only for top squarks. \s 

The total cross sections for the associated production of the lightest
CP--even Higgs boson, $\ee \ra h\gamma$, in the unpolarized case are
shown in Figs.~4--7. The cross sections for longitudinally polarized
electron and positron beams will not be displayed: as in the SM case, 
they can be obtained by simply rescaling the figures by approximately factor 2 
for left--handed electrons and a factor 4 for left-handed electrons
and right-handed positrons. \s 

Fig.~4 shows the unpolarized cross sections at a centre of mass energy
$\sqrt{s}=500$ GeV as a function of $M_h$ for two values of $\tb$, $\tb=2$ 
and $50$. In Fig.~4a(4b), the SUSY parameters have been chosen
in such a way that charginos, neutralinos and sleptons will (not) be
kinematically reachable at a 500 GeV $\ee$ collider. The solid
lines show the total cross sections including the SUSY contributions,
while the dashed lines include only the standard and Higgs contributions
as it is the case for a two--Higgs doublet model (2HDM) where the MSSM 
relations have been implemented. \s

For low $M_h$ the cross sections are smaller than in the SM, especially
for large $\tb$ values. This is due to the fact that the main
contributions are coming from the $W$ bosons loops, and in the MSSM or
2HDM, the $W$ boson couplings are suppressed by a factor $\sin(\beta-\alpha)$
compared to the SM Higgs coupling. The suppression can be rather
drastic, especially for large values of $\tb$, where the cross sections
drop to very small values. This can be best seen in Fig.~5 where the
cross sections are plotted against $\tb$ for two values of $M_A=50$ and
$500$ GeV. With increasing $M_h$, the cross sections increase and in
the 2HDM they reach, as it should be, the SM cross section values in the
decoupling limit where $M_h$ is maximal [in practice for $M_A \sim 500$
GeV which leads to $M_h \sim 90$ GeV for $\tb=2$ and $M_h \sim 120$ GeV
for $\tb=50$]. \s 

The contributions of the SUSY particles interfere destructively with the
contributions of the standard particles in most of the parameter space.
The only exception is when the standard contributions give a very small
cross section, as in the case of very high $\tb$ values far from the
decoupling limit. In the scenario of Fig.~4a and Fig.~5a, where SUSY
particles can be found directly at a c.m. energy of 500 GeV, the
difference between the MSSM and the 2HDM is very large, more than a
factor of 2. Unfortunately, since the cross section is smaller than in 
the SM, the signal will be harder to isolate compared to the case of the 
standard Higgs boson. For the scenario where the SUSY particles are too heavy 
to be
produced directly at the chosen c.m. energy, the SUSY contributions are
very small and they will be very difficult to be detected. \s 

This is best illustrated in Fig.~6, where the relative difference
between the SM and the MSSM cases in the decoupling limit $M_A \gg M_Z$.
In this limit $h$ and the standard $H^0$ have practically the same
couplings and it is very hard to distinguish between the SM and the
MSSM. This difference is plotted against $M_2$ in Fig.~6a and $\mu$ in
Fig.~6b, for $\tb=2,50$ and for several values of $\mu$ and $M_2$,
respectively. For small values of the parameters $M_2$ or $\mu$,
charginos are light enough to give sizeable contributions, especially
near the production threshold. For larger values of these parameters,
the SUSY particles are too heavy and since their couplings are not
proportional to their masses, they decouple from the production
amplitudes rendering the difference between the MSSM and the SM
contributions very small. \s 

In Fig.~7, the cross section $\ee \ra h\gamma$ is shown as a function of
the c.m. energy for $\tb=2,50$ and $M_A=50$ and 500 GeV. As expected
from the SM case, the cross sections decrease smoothly [except when some
threshold for the real production of a particle is crossed] with
increasing $\sqrt{s}$. Therefore, to optimize the cross section, one
needs to operate the collider at energies around $\sqrt{s}=200$ GeV.
This is about the LEP2 c.m. energy; but at LEP2, the luminosity is so
small that no event can be expected. \s 

Finally, Figs.~8 and 9 display the cross sections for the associated
production of the heavy CP--even Higgs boson $H$ [Fig.~8] and the
pseudoscalar Higgs particle $A$ [Fig.~9] at $\sqrt{s}=500$ GeV as a
function of the Higgs masses. The same scenarios as in Fig.~4 have been
chosen. Except for light $H$ and $A$ and for small values of $\tan
\beta$, the cross sections are very small. Again, the difference between
the MSSM and 2HDM is large only when the SUSY particles are within the
reach of the collider.

\section*{4. Conclusions}

We have calculated the cross sections for the production of Higgs
particles in association with a photon in $\ee$ collisions, allowing for
the longitudinal polarization of the initial $e^-$ and $e^+$ beams. We
have considered the case of the Standard Model Higgs boson, $\ee \ra
\gamma H^0$, and the case of the neutral CP--even and CP--odd Higgs
particles of the Minimal Supersymmetric extension of the Standard Model,
$\ee \ra \gamma h$, $\gamma H$ and $\gamma A$. We have given complete
and compact analytical expressions, and made careful checks of the gauge
invariance of the results, as well as a comparison with earlier results
for the associated production of the SM Higgs boson in the case of
unpolarized beams. \s 

We have then illustrated the  size of the cross sections for energies
which will be reached at future $\ee$ colliders. In the SM, the cross
sections are in general small but the signals are rather clean. With an
integrated luminosity of ${\cal L} \sim 100$ fb$^{-1}$ expected at
future high--energy $\ee$ colliders and with the polarization of both
electron and positron beams, which increase the cross sections by a
factor of 4 compared to the unpolarized case, one could hope to isolate
the signals despite of the low rates. \s 

The cross sections for the associated production of the lightest MSSM
CP--even Higgs boson are in general smaller than for the SM Higgs boson
except in the decoupling limit where they reach the SM values. The
contribution of the SUSY particles are significant only when these
particles are light enough to be produced directly at the collider. For
SUSY particles beyond the reach of the $\ee$ collider, the difference
between the SM and MSSM cross sections is very small and will be hard to
be detected. For the heavy CP--even and the CP--odd Higgs bosons, the cross
sections are rather small especially for large values of $\tb$. Only for
small values of $\tb$ and relatively small $A$ and $H$ masses that the
cross sections exceed $10^{-2}$ fb in the unpolarized case. 

\newpage

\subsection*{APPENDIX: Scalar Functions}

\setcounter{equation}{0}
\renewcommand{\theequation}{A.\arabic{equation}}

Our conventions for the scalar Passarino--Veltman functions and for
the tensor integrals are as follows [$\mu$ is the renormalization 
scale and $D$ the space--time dimension] 
\beq
\frac{i}{16\pi^2} \ A(m^2)  \equiv \mu^{4-D} \int \frac{d^D k} 
{(2\pi)^D} \frac{1}{k^2-m^2}  
\eeq
\beq
\frac{i}{16\pi^2} \ B_{0}(p_1^2;m_1^2,m_2^2) \equiv 
\mu^{4-D} \int \frac{d^D k} {(2\pi)^D} 
\frac{1} {[k^2-m_1^2][(k+p_1)^2-m_2^2]}  
\eeq
\beq
\frac{i}{16\pi^2} && C_{0;\mu;\mu\nu}(p_1^2,p_2^2,p_3^2;m_1^2,m_2^2,m_3^2)
\equiv \mu^{4-D}  \non \\
&& \int \frac{d^D k}{(2\pi)^D} \frac{1;k_{\mu};k_{\mu}k_{\nu} }
 {[k^2-m_1^2][(k+p_1)^2-m_2^2][(k+p_1+p_2)^2-m_2^3]} 
\eeq
\beq
\frac{i}{16\pi^2} D_{0;\mu;\mu\nu}
(p_1^2,p_2^2,p_3^2,p_4^2,(p_1+p_2)^2,(p_2+p_3)^2;m_1^2,m_2^2,m_3^2,m_4^2) 
\equiv  \mu^{4-D} \non \\
\int \frac{d^D k}{(2\pi)^D} 
\frac{1;k_{\mu};k_{\mu}k_{\nu}}
 {[k^2-m_1^2][(k+p_1)^2-m_2^2][(k+p_1+p_2)^2-m_3^2][(k+p_1+p_2+p_3)^2-m_4^2]} 
 \non \\
\eeq
with the tensor decomposition
\beq
% B^{\mu} &=& p_1^{\mu} B_{1} \\
% B^{\mu\nu} &=&
%      p_1^{\mu} p_1^{\nu} B_{21} + g^{\mu\nu} B_{22} \\
C^{\mu} &=& p_1^{\mu} C_{11}+ p_2^{\mu} C_{12} 
\\
C^{\mu\nu} &=&
     p_1^{\mu} p_1^{\nu} C_{21} + p_2^{\mu} p_2^{\nu} C_{22}
   +( p_1^{\mu} p_2^{\nu}+p_1^{\nu} p_2^{\mu} ) C_{23} + g^{\mu\nu} C_{24}
\\
D^{\mu} &=& p_1^{\mu} D_{11} + p_2^{\mu} D_{12} + p_3^{\mu} D_{13} 
 \\
D^{\mu\nu} &=&
     p_1^{\mu} p_1^{\nu} D_{21}
   + p_2^{\mu} p_2^{\nu} D_{22}
   + p_3^{\mu} p_3^{\nu} D_{23} \non\\
&& +( p_1^{\mu} p_2^{\nu}+p_1^{\nu} p_2^{\mu} ) D_{24}
   +( p_1^{\mu} p_3^{\nu}+p_1^{\nu} p_3^{\mu} ) D_{25}
   +( p_2^{\mu} p_3^{\nu}+p_2^{\nu} p_3^{\mu} ) D_{26}
   + g^{\mu\nu} D_{27} \non \\
\eeq

The analytical expressions of the scalar functions can be found e.g. 
in Ref.~\cite{S3}. 

\newpage

\subsection*{ Table 1:}

\setlength{\tabcolsep}{4.mm}
\renewcommand{\baselinestretch}{1.2}
\renewcommand{\arraystretch}{1.2}

\begin{center}
\begin{tabular}{|c||c|c|c|} \hline\rule[0mm]{0mm}{7mm} 
$\hspace{.5cm}g/ \Phi \hspace{.5cm}$ &  h & H & \hspace{.5cm}A\hspace{.5cm} 
\\[0.2cm] \hline \hline 
$g_{\Phi VV} $ &  $ \sin(\beta-\alpha) $  &  $ \cos (\beta-\alpha)  $ & 0 
\\[0.1cm] \hline \hline
$g_{\Phi AZ} $ &  $ \cos (\beta-\alpha) $ &  $  - \sin (\beta-\alpha)  $ & 0
\\[0.1cm] \hline 
$g_{\Phi H^\pm W^\pm} $ & $ \mp \cos (\beta-\alpha) $ &  
$ \pm \sin (\beta-\alpha)  $  & 1
\\[0.1cm] \hline \hline
$g_{\Phi G^+G^-} $ &  $ \cos(2\beta)\sin(\beta+\alpha) $  
&  $ \cos(2\beta)\cos (\beta+\alpha)  $ & 0 
\\[0.1cm] \hline 
\end{tabular}
\end{center}
%\vspace*{2mm}

\nn {\small {\bf Tab.~1a}: The Higgs--vector boson couplings $g_{\Phi VV}$
[normalized to the SM Higgs coupling $g_{H^0VV}=2 \left[
\sqrt{2}G_F\right]^{1/2} M_V^2$], and the Higgs--Higgs--vector boson
couplings [normalized to $g_W=( \sqrt{2} G_F)^{1/2}$ $M_W$ and
$g_Z=(\sqrt{2}G_F)^{1/2}M_Z$ for the charged/neutral weak couplings];
the latter come with the sum of the Higgs momenta entering and leaving
the vertices.}

\vspace*{5mm}

\begin{center}
\begin{tabular}{|c||c|c|} \hline
\rule[0mm]{0mm}{7mm} 
$ \hspace{1cm} \Phi \hspace{1cm} $ &$ g_{ \Phi \bar{u} u} $ & $
g_{\Phi \bar{d} d} $ \\[2mm]
 \hline \hline
\rule[0mm]{0mm}{7mm} 
$h$  & \ $\; \cos\alpha/\sin\beta       \; $ \ & \ $ \; -\sin\alpha/
\cos\beta \; $  \\
$H$  & \       $\; \sin\alpha/\sin\beta \; $ \ & \ $ \; \cos\alpha/
\cos\beta \; $  \\
$A$  & \ $\; 1/ \tb \; $        \ & \ $ \; \tb \; $  \\[0.3cm] \hline
\end{tabular}
\end{center}

\vspace*{3mm}

\nn {\small {\bf Tab.~1b}: Higgs boson couplings in the MSSM to fermions 
and gauge bosons relative to the SM Higgs  couplings.}

\bigskip
\begin{center}
\begin{tabular}{|c||c|c|c|} \hline\rule[0mm]{0mm}{7mm} 
$g^{L,R} / \Phi $
&  $h$  &  $H$  &  $A$  \\[0.2cm] \hline \hline 
$g^L_{\Phi \chi^+_i \chi_j^-}$ &
$Q^*_{ji}\sin\alpha - S^*_{ji}\cos\alpha $ &
$-Q^*_{ji}\cos\alpha - S^*_{ji}\sin\alpha $ &
$-Q^*_{ji}\sin\beta - S^*_{ji}\cos\beta$  \\
$g^R_{\Phi \chi^+_i \chi_j^-}$ &
$Q_{ij}\sin\alpha - S_{ij}\cos\alpha $ &
$-Q_{ij}\cos\alpha - S_{ij}\sin\alpha $ &
$Q_{ij}\sin\beta + S_{ij}\cos\beta$
\\[2mm] \hline
$g^L_{\Phi \chi^0_i \chi_j^0}$ &
$Q^{"*}_{ji}\sin\alpha + S^{"*}_{ji}\cos\alpha $ &
$-Q^{"*}_{ji}\cos\alpha + S^{"*}_{ji}\sin\alpha $ &
$-Q^{"*}_{ji}\sin\beta + S^{"*}_{ji}\cos\beta$  \\
$g^R_{\Phi \chi^0_i \chi_j^0}$ &
$Q^"_{ij}\sin\alpha + S^"_{ij}\cos\alpha $ &
$-Q^"_{ij}\cos\alpha + S^"_{ij}\sin\alpha $ &
$Q^"_{ij}\sin\beta - S^"_{ij}\cos\beta$  \\[2mm] \hline
\end{tabular}
\end{center}
%\vspace*{2mm}

\nn {\small {\bf Tab.~1c}: The couplings of the neutral Higgs bosons to
charginos and neutralinos, normalized to $g_{W}=
\left[\sqrt{2}G_F\right]^{ 1/2} M_W$; the matrix elements $Q_{ij}/
S_{ij}$ and $Q^{"}_{ij}/S^{"}_{ij}$ can be found in Ref.~\cite{S5}.}

\newpage

\subsection*{ Table 2:}

\vspace*{1cm}

\begin{center}
\begin{tabular}{|c|c|c|c|c|c|} \hline\rule[0mm]{0mm}{7mm} 
$ \ \ \tilde{f} \ \ $ & $\ \ \Phi \ \ $ & $\ \ g_1^{\Phi } \ \ $ 
& $g_2^{\Phi }$ & $g_3^{\Phi }$ & $g_4^{\Phi } $  \\[2mm] \hline
 \rule[0mm]{0mm}{7mm} 
&$h$ & $\cos\alpha/ \sin \beta $ & $-\sin(\alpha+\beta)$ 
               & $-\sin\alpha/\sin\beta$ & $\cos \alpha/ \sin\beta$ \\
$ \tilde{u}$& $H$ & $\sin\alpha/\sin\beta$ & $\cos(\alpha+\beta)$ & 
    $\cos \alpha /\sin \beta$ & $\sin \alpha/ \sin\beta$ \\ 
& $A$ & $0$ & $0$ & $1$ & $-1/\tb$ \\[3mm]
& $h$ & $-\sin\alpha/ \cos \beta $ & $-\sin(\alpha+\beta)$ 
               & $\cos\alpha/\cos\beta$ & $-\sin \alpha/ \cos\beta$ \\
$\tilde{d}$ & $H$ & $\cos\alpha/\cos\beta$ & $\cos(\alpha+\beta)$ & 
    $\sin \alpha /\cos \beta$ & $\cos \alpha/ \cos\beta$ \\ 
&    $A$ & $0$ & $0$ & $1$ & $-\tb$ \\[2mm] \hline
\end{tabular}
\end{center}

\nn {\small {\bf Tab.~2a}: Coefficients in the couplings of 
neutral Higgs bosons to sfermion pairs. } 

\vspace*{1cm}

\bigskip
\begin{center}
\begin{tabular}{|c||c|c|c|} \hline \rule[0mm]{0mm}{7mm} 
$\hspace{.5cm} \tilde{l}_i \tilde{l}_j \hspace{.5cm}$
&  $g_{h \tilde{l}_i \tilde{l}_j} $
&  $g_{H \tilde{l}_i \tilde{l}_j} $
&  $\ \ g_{A \tilde{l}_i \tilde{l}_j} \ \ $ \\[0.2cm] \hline \hline
\rule[0mm]{0mm}{7mm} 
$\tilde{e}_L \tilde{e}_L $
& $ \ (2s_W^2-1) \sin(\beta+\alpha) \ $
& $ \ -(2s_W^2-1) \cos(\beta+\alpha) \ $
& $ 0$ \\
$\tilde{e}_R \tilde{e}_R $
& $ \ 2s_W^2 \sin(\beta+\alpha) \ $
& $ \ -2s_W^2  \cos(\beta+\alpha) \ $
& $ 0$ \\
$\tilde{\nu}_L \tilde{\nu}_L $
& $ \ \sin(\beta+\alpha) \ $
& $ \ - \cos(\beta+\alpha) \ $
& $ 0$ \\[2mm] \hline
\end{tabular}
\end{center}
%\vspace*{2mm}

\nn {\small {\bf Tab.~2b}: The couplings of the neutral Higgs bosons to left--
and right--handed sleptons in the absence of mixing, normalized to $g_{W}'= 
\left[\sqrt{2}G_F\right]^{1/2} M_W^2$.}

\newpage

%
% -------------------------------------------------------------
\def\npb#1#2#3{{\rm Nucl. Phys. }{\rm B #1} (#2) #3}
\def\plb#1#2#3{{\rm Phys. Lett. }{\rm #1 B} (#2) #3}
\def\prd#1#2#3{{\rm Phys. Rev. }{\rm D #1} (#2) #3}
\def\prl#1#2#3{{\rm Phys. Rev. Lett. }{\rm #1} (#2) #3}
\def\prc#1#2#3{{\rm Phys. Rep. }{\rm C #1} (#2) #3}
\def\pr#1#2#3{{\rm Phys. Rep. }{\rm #1} (#2) #3}
\def\zpc#1#2#3{{\rm Z. Phys. }{\rm C #1} (#2) #3}
\def\nca#1#2#3{{\it Nouvo~Cim.~}{\bf #1A} (#2) #3}
%
% **************************  References ****************************

\newpage

\nn {\Large \bf Figure Captions}

\begin{itemize}

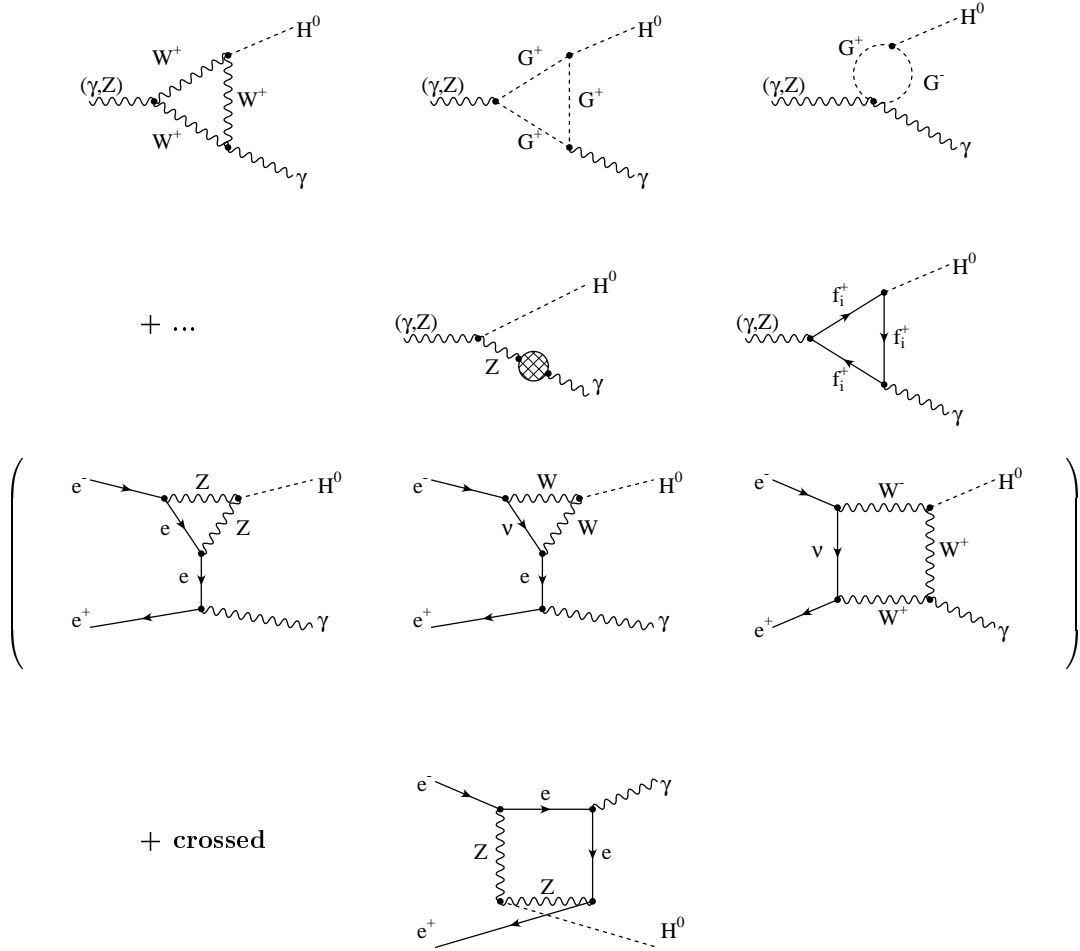
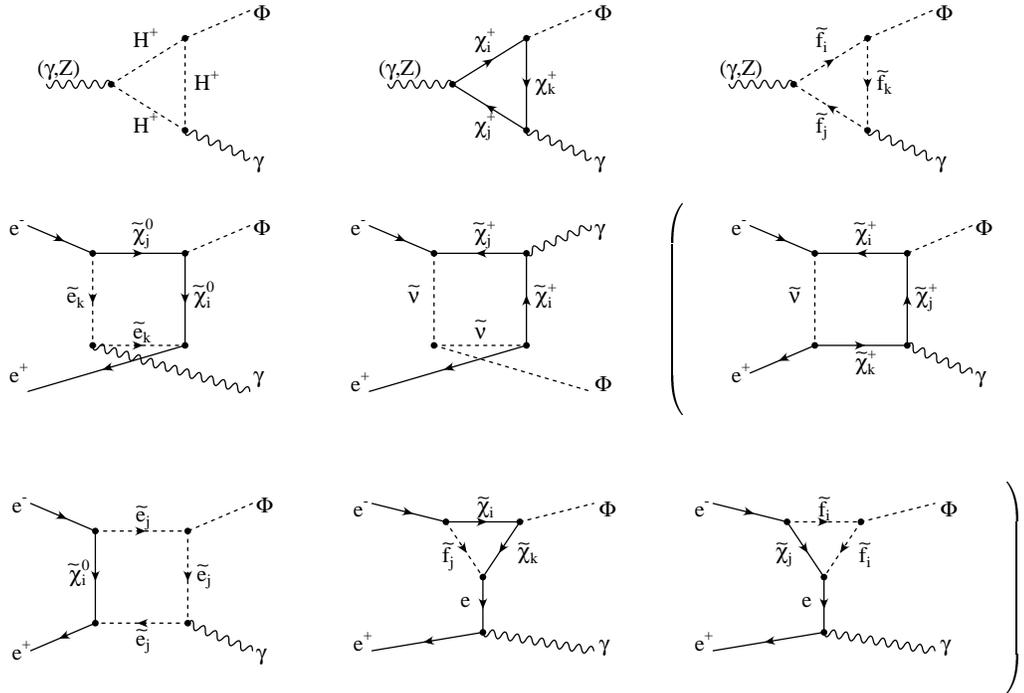
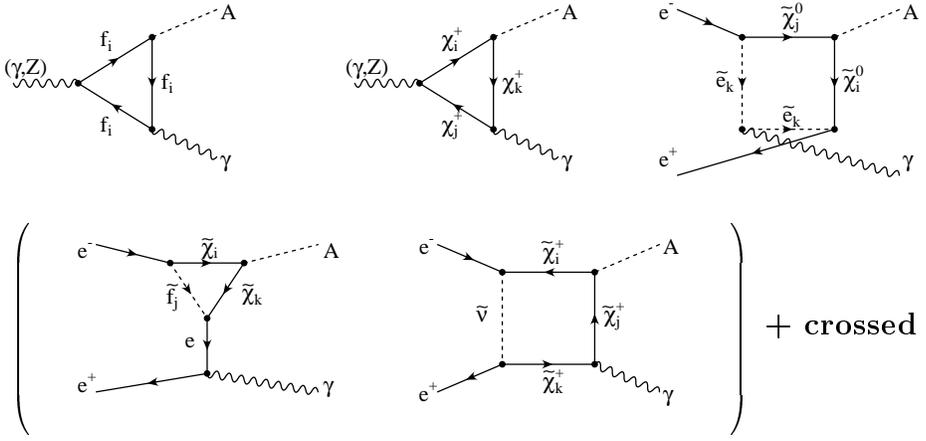
\item[{\bf Fig.~1:}] 
Feynman diagrams contributing to the process $\ee \ra \gamma H^0$ in
the Standard Model (a), the additional diagrams for the production of
the MSSM CP--even Higgs bosons $\ee \ra \Phi \gamma$ with $\Phi=h,H$
(b), and of the CP--odd Higgs boson $\ee \ra A \gamma$ (c). 

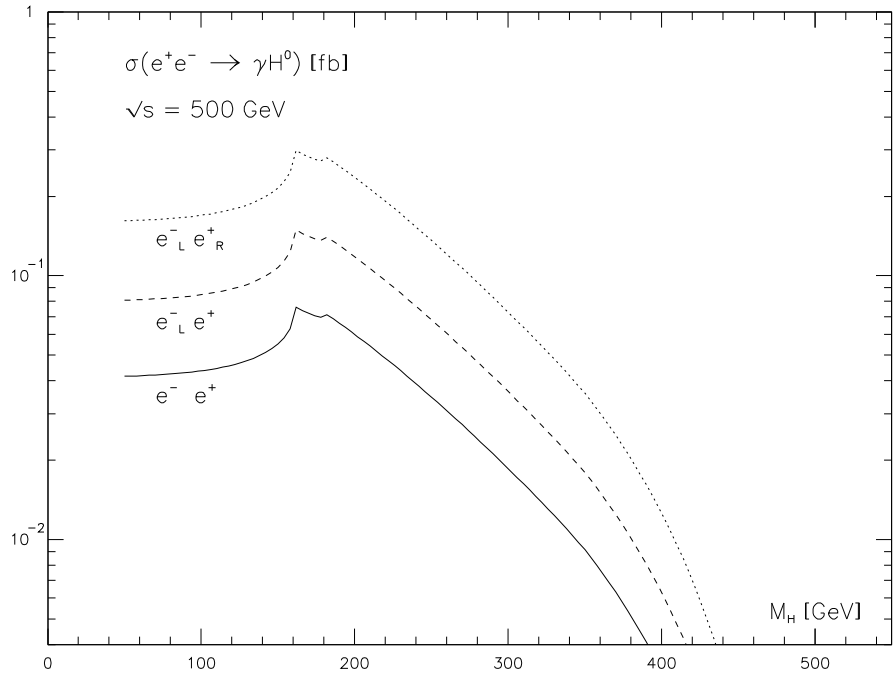
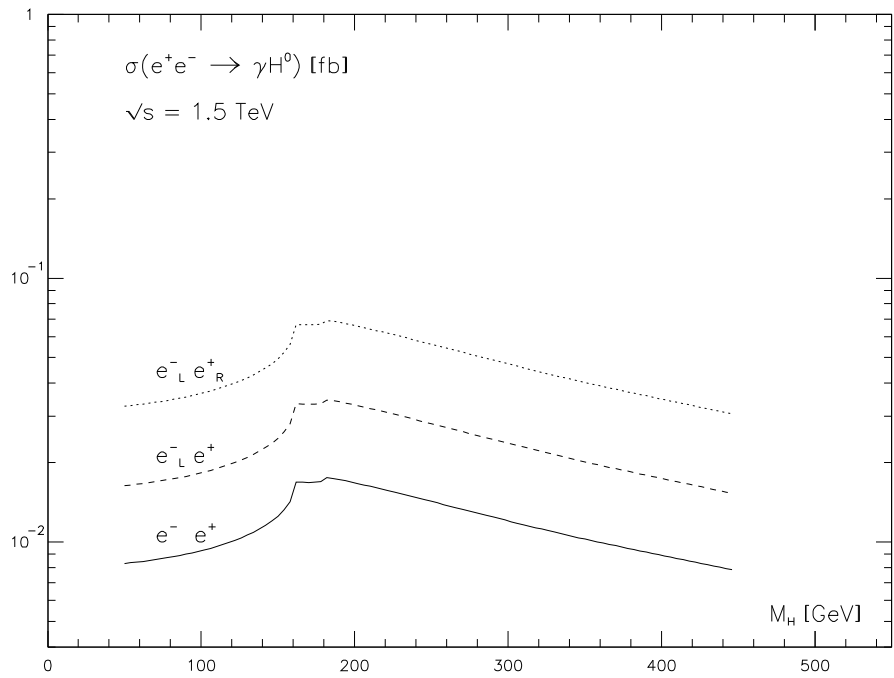
\item[{\bf Fig.~2:}] 
The total cross sections for the production of the Standard Model
Higgs boson, $e^+e^- \to \gamma H^0$, at two center of mass energies 
$\sqrt{s} = 500 $ GeV and $\sqrt{s} = 1.5$ TeV as a function of the
Higgs boson mass. The solid, dashed  and dotted lines are for the 
unpolarized, left--handed polarized electrons, and left--handed 
polarized electrons and right--handed polarized positrons, respectively. 

\item[{\bf Fig.~3:}] 
(a) The total production cross section $\sigma(e^+e^- \to \gamma H^0)$
for unpolarized beams as a function of $\sqrt{s}$ for $M_H=$ 100 (solid), 
150 (dot--dashed), 200 (dashed) and $300$ (dotted) GeV. 
(b) The differential cross section $d\sigma/d\cos\theta$ at a center 
of mass energy $\sqrt{s} = 500$ GeV for $M_H = 100$ GeV for unpolarized and
polarized $\ee$ beams. 

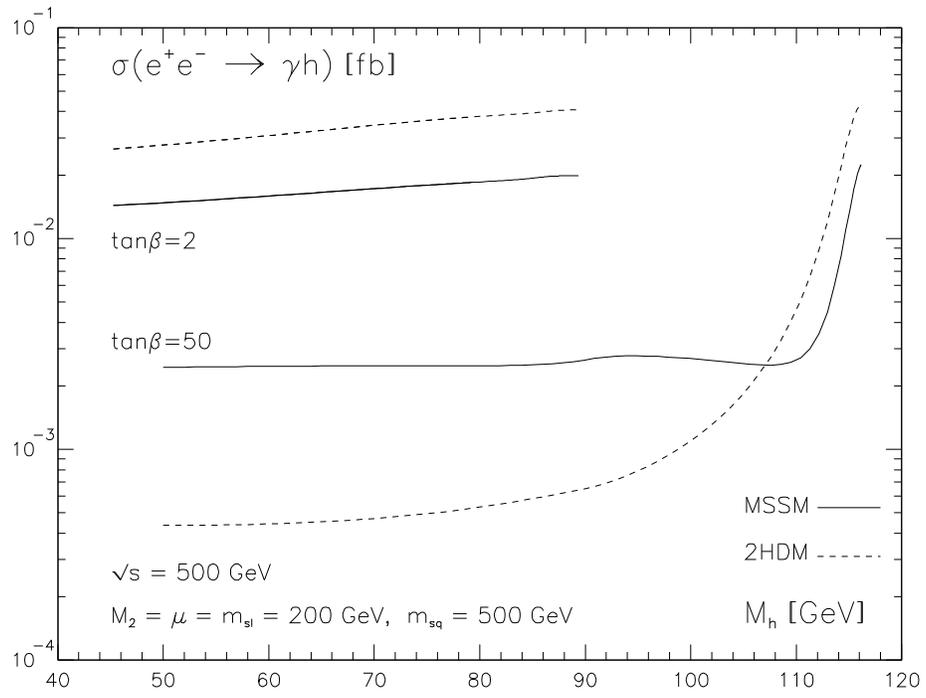
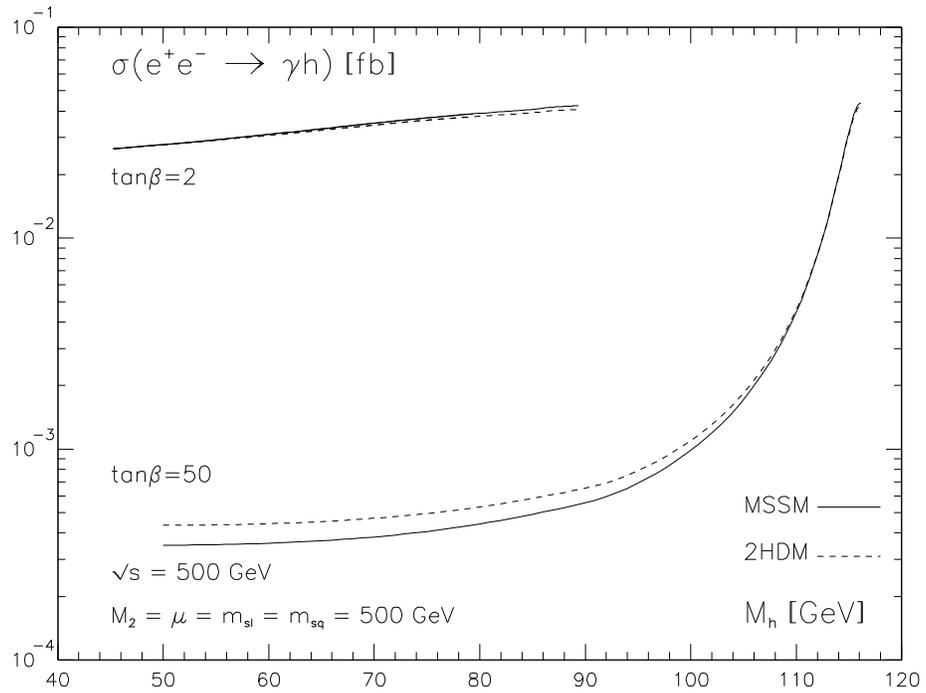
\item[{\bf Fig.~4:}] 
The total cross section for the production of the lightest
MSSM Higgs boson, $\sigma(e^+e^- \to \gamma h)$, in the
unpolarized case at $\sqrt{s} = 500$ GeV as a function of $M_h$ for
$\tan\beta=2$ and $50$.  The solid lines include all MSSM contributions,
while the dashed lines include only the contributions from the standard
and Higgs sectors (2HDM). For the SUSY parameters we have: in  (a)
$M_2 = 200$ GeV, $\mu = 200$ GeV, $M_{\tilde{l}} = 200$ GeV, 
$M_{\tilde{q}} = 500$ GeV, and in (b) $M_2 = 500$ GeV, $\mu = 500$ GeV,
$M_{\tilde{l}} = 500$ GeV, $M_{\tilde{q}} = 500$ GeV.

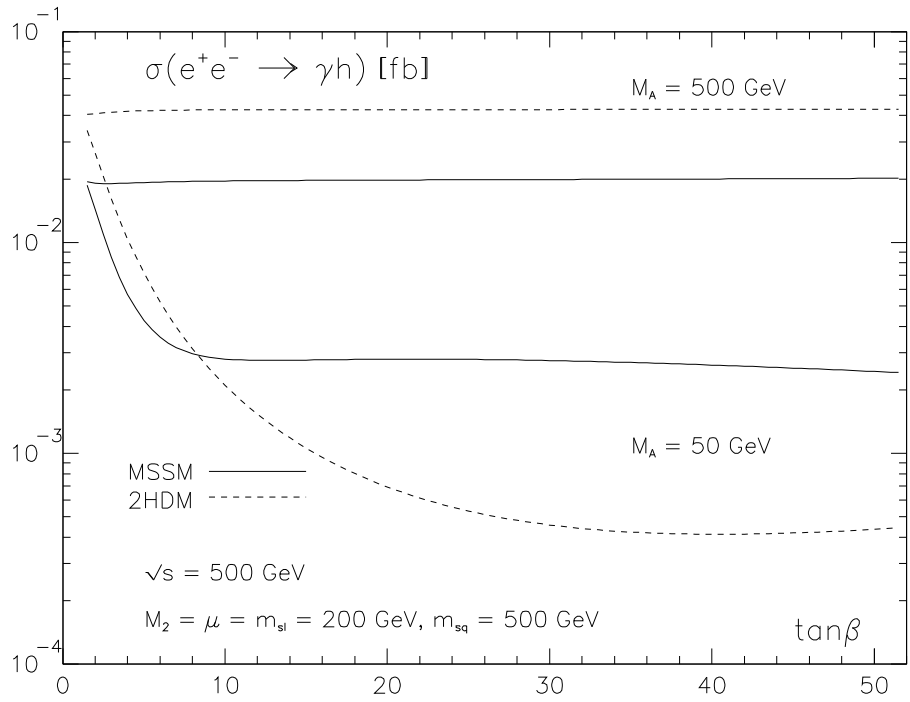
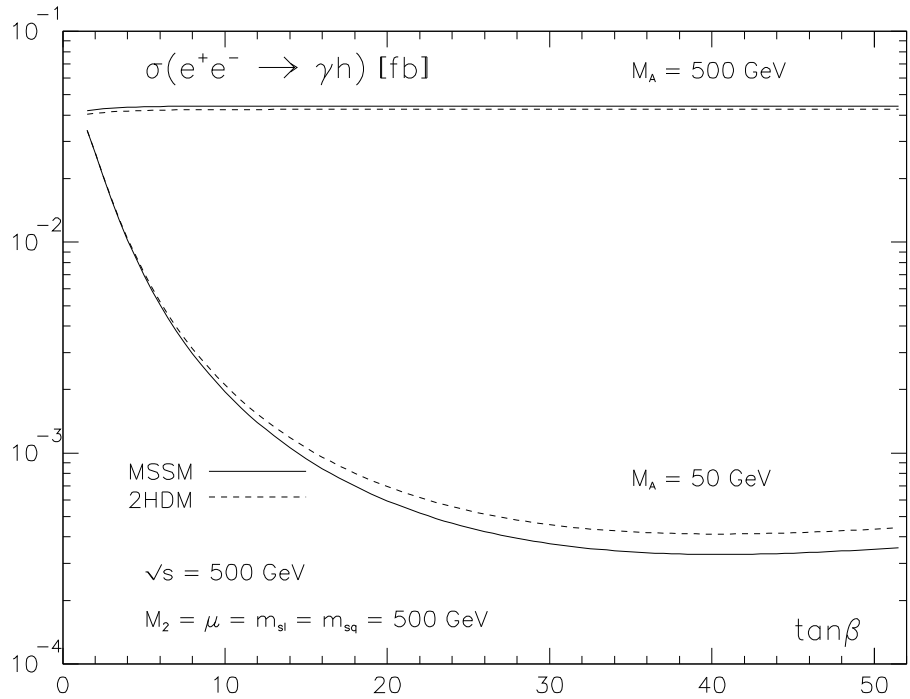
\item[{\bf Fig.~5:}] 
The total production cross section $\sigma(e^+e^- \to \gamma h)$ for
unpolarized beams at $\sqrt{s} = 500$ GeV as a function of $\tan\beta$
for  $M_A = 50$ GeV and $M_A=500$ GeV. The scenarios for the MSSM
parameter space are as in Fig.~4. 

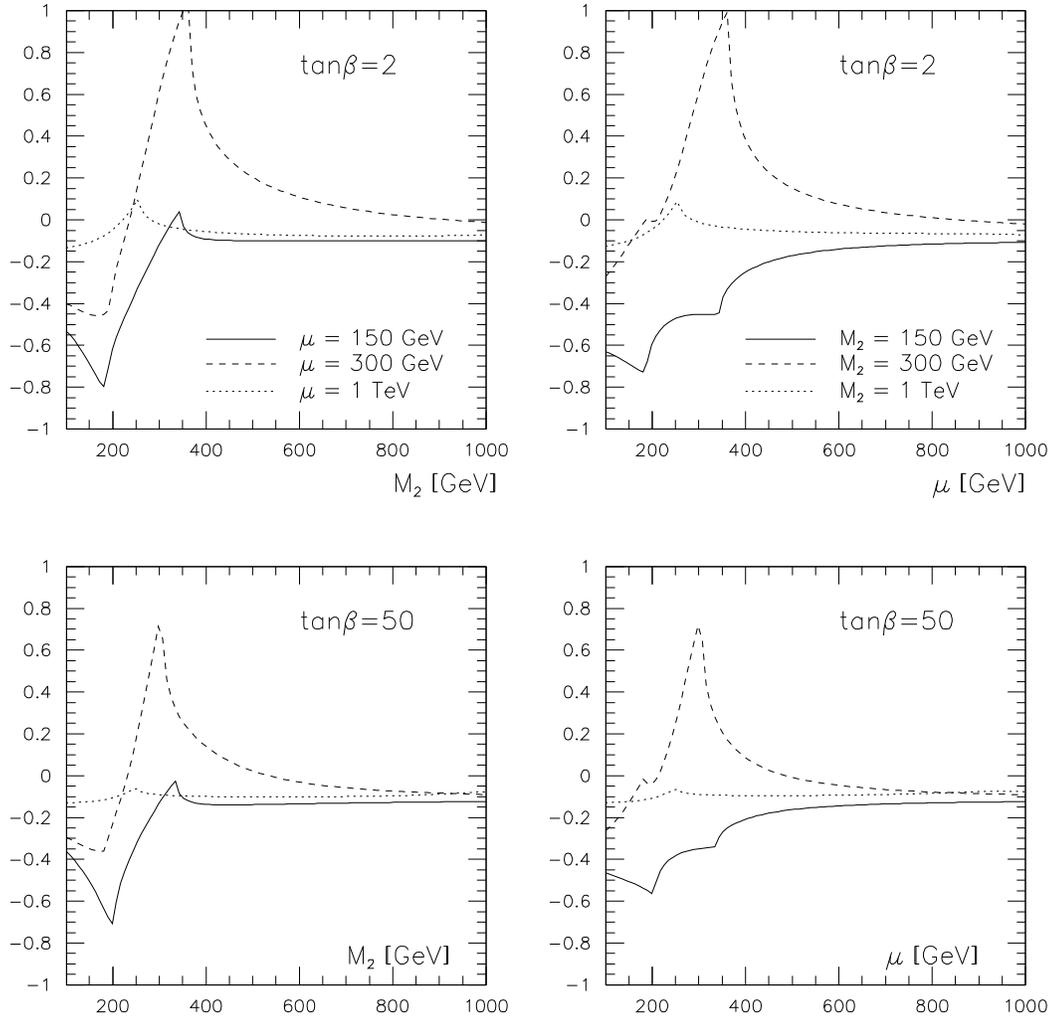
\item[{\bf Fig.~6:}] 
The difference between the MSSM and the SM cross sections at $\cos\theta 
= 0$ normalized to the SM values, as a function of the SUSY parameters 
$M_2$ and $\mu$ for $\tan\beta=2$ and $50$. For the other parameters, we 
take $M_A = 1$ TeV and $M_{\tilde{q}} = M_{\tilde{l}} = 500$ GeV.

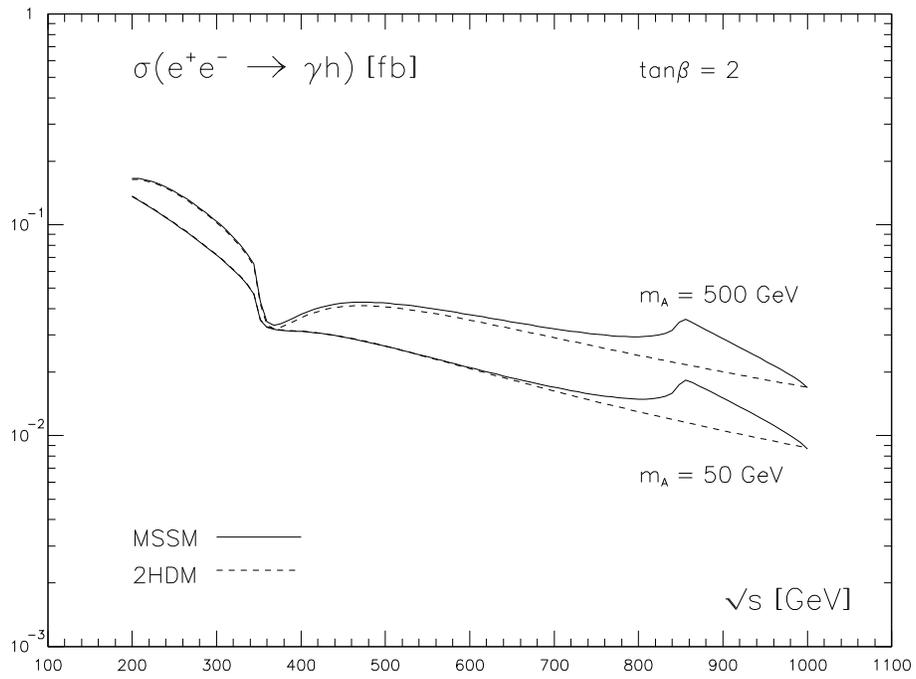
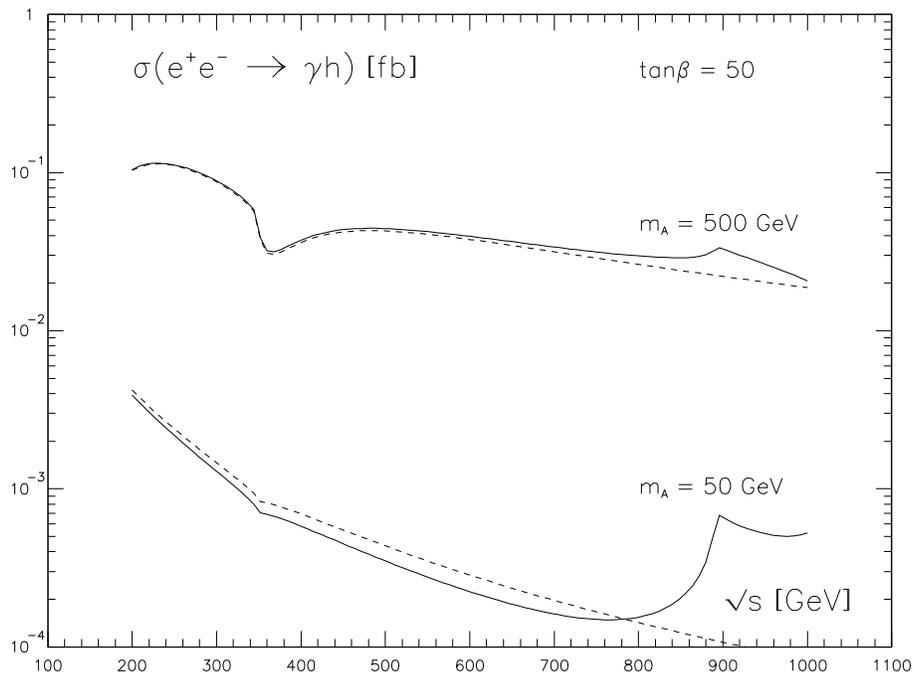
\item[{\bf Fig.~7:}] 
The total production cross section $\sigma(e^+e^- \to \gamma h)$ as a
function of the c.m. energy $\sqrt{s}$ for $M_A = 50$ GeV and $M_A500$
GeV and $\tan\beta=2$ and $50$ in the MSSM (solid curves) and in the
2HDM (dashed curves). The SUSY parameters are chosen as: $M_2 = 500$
GeV, $\mu = 500$ GeV, $M_{\tilde{l}} = 500$ GeV and  $M_{\tilde{q}} =
500$ GeV.

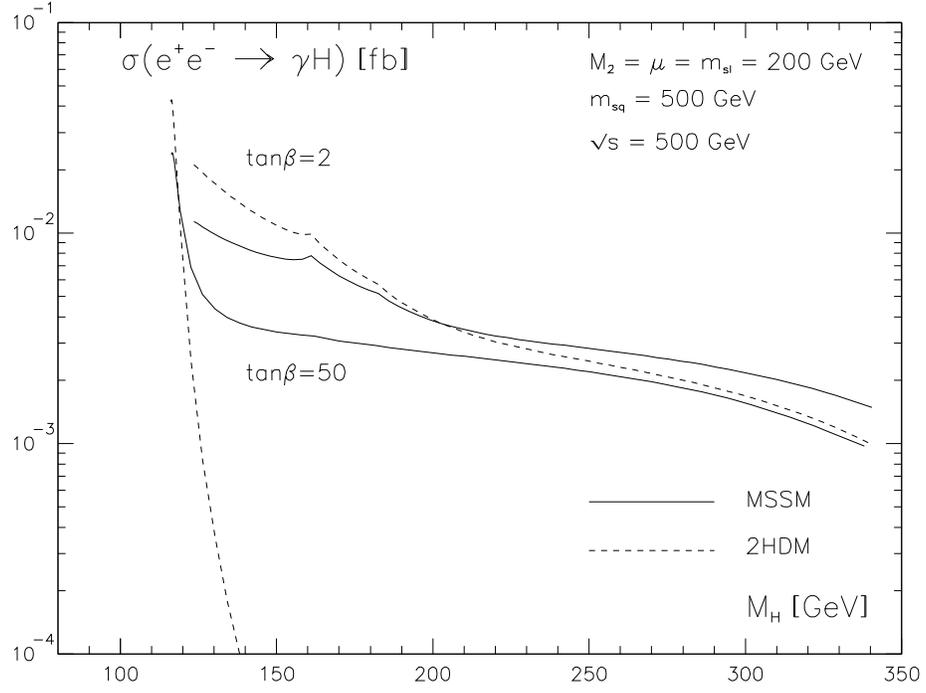
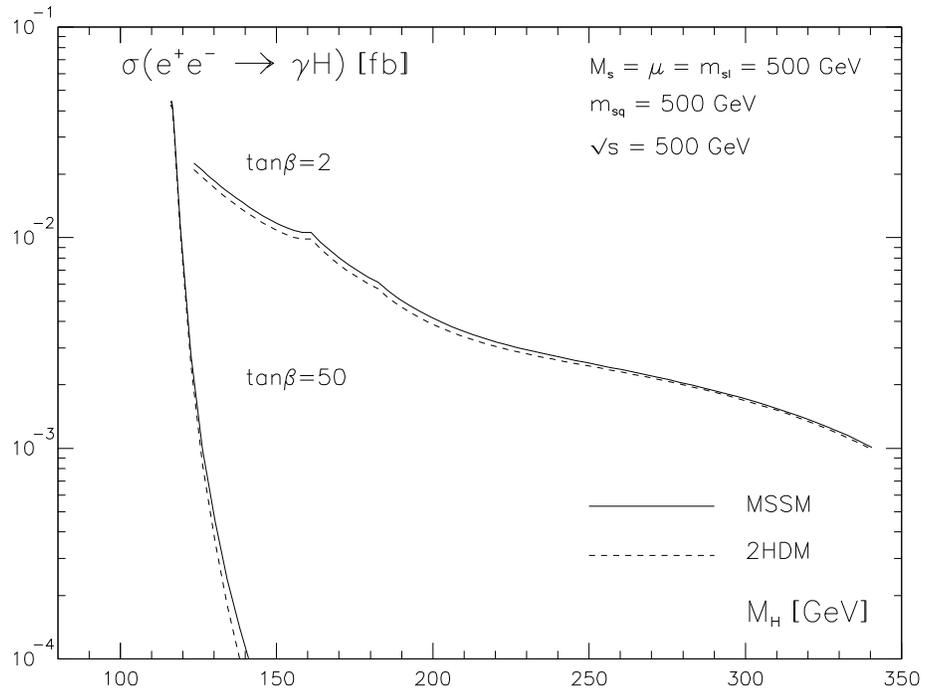
\item[{\bf Fig.~8:}] 
The total cross section for the production of the heavy CP--even Higgs
boson in the MSSM, $\sigma(e^+e^- \to \gamma H)$, 
in the unpolarized case at $\sqrt{s} = 500$ GeV as a function of $M_H$ 
for $\tan\beta=2$ and $50$ in the MSSM (solid curves) and 2HDM (dashed
curves). The SUSY parameters are as in Fig.~4. 

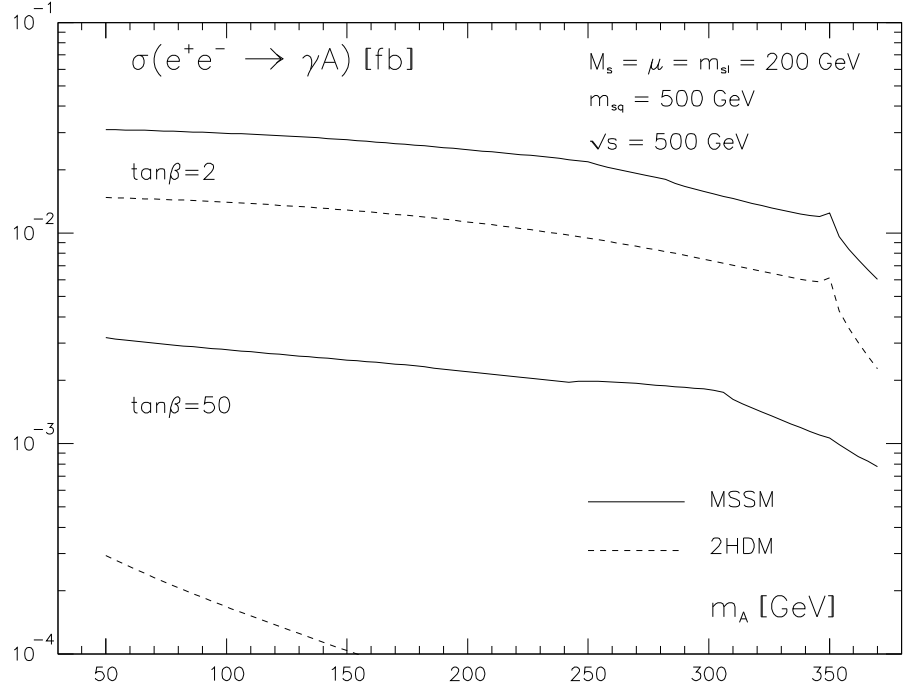
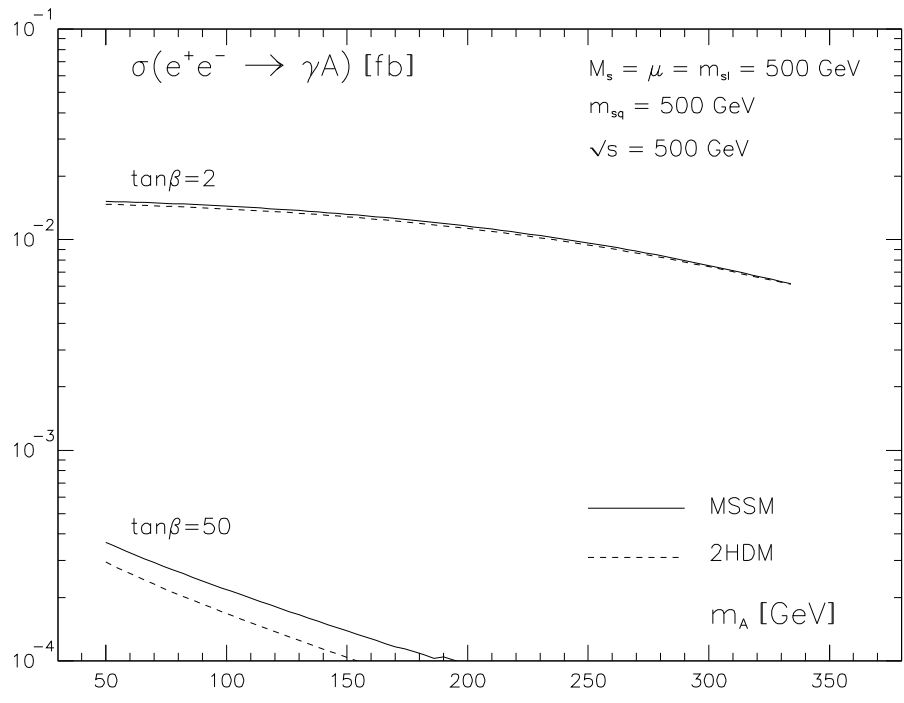
\item[{\bf Fig.~9:}] 
  The total cross section for the production of the CP--odd Higgs
boson in the MSSM, $\sigma(e^+e^- \to \gamma A)$, 
in the unpolarized case,  at $\sqrt{s} = 500$ GeV as a function of $M_A$ 
for $\tan\beta=2$ and $50$ in the MSSM (solid) and 2HDM (dashed). 
The SUSY 
parameters are as in Fig.~4. 

\end{itemize}

%%%%%%%%%%%%%%%%%%%%%%       

\end{document}